\documentclass[aps,prb,twocolumn,noeprint]{revtex4-1}
\usepackage{graphicx}
\usepackage{amsmath,amsfonts,amsthm,bm}
\usepackage{tikz}
\usetikzlibrary{decorations.pathreplacing}
\usepackage{subcaption}
\usepackage[bookmarks=false]{hyperref}
\hypersetup{colorlinks=true, citecolor=blue, urlcolor=blue, linkcolor=blue}

\makeatletter
\renewcommand{\maketag@@@}[1]{\hbox{\m@th\normalsize\normalfont#1}}%
\makeatother

\begin{document}

\title{Fluctuating-time and full counting statistics for quantum transport in a system with internal telegraphic noise}
\author{Samuel L. Rudge}
\author{Daniel S. Kosov}
\address{College of Science and Engineering, James Cook University, Townsville, QLD, 4814, Australia }

\begin{abstract}
Many molecular junctions display stochastic telegraphic switching between two distinct current values, which is an important source of fluctuations in nanoscale quantum transport. Using Markovian master equations, we investigate electronic fluctuations and identify regions of non-renewal behavior arising from telegraphic switching. Non-renewal behavior is characterized by the emergence of correlations between successive first-passage times of detection in one of the leads. Our method of including telegraphic switching is general for any source-molecule-drain setup, but we consider three specific cases. In the first, we model stochastic transitions between an Anderson impurity with and without an applied magnetic field $B$. The other two scenarios couple the electronic level to a single vibrational mode via the Holstein model. We then stochastically switch between two vibrational conformations, with different electron-phonon coupling $\lambda$ and vibrational frequency $\omega$, which corresponds to different molecular conformations. Finally, we model the molecule attaching and detaching from an electrode by switching between two different molecule-electrode coupling strengths $\gamma$. We find, for all three cases, that including the telegraph process in the master equation induces relatively strong positive correlations between successive first-passage times, with Pearson coefficient $p \approx 0.5$. These correlations only appear, however, when there is telegraphic switching between two significantly different transport scenarios, and we show that it arises from the underlying physics of the model. We also find that, in order for correlations to appear, the switching rate $\nu$ must be much smaller than $\gamma$.
\end{abstract}

\maketitle

\section{Introduction}

The physical differences between nanoscale and macroscopic conductors are best exemplified by the presence of electronic fluctuations, which universally occur in the former yet rarely occur in the latter. Fluctuations arise in nanoscale systems from various sources: the unavoidable probabilistic nature of quantum transport, discrete charge carriers coupled with low currents, and stochastic changes in intra-system dynamics \cite{Nazarov2009}. We are interested in this last source, when the electric current stochastically moves between two different values: commonly referred to as telegraphic switching or a telegraph process.

Telegraphic switching is a common experimental phenomenon, which is visible in scenarios containing two or more distinct states with different parameters governing the transport. Telegraph noise, which is distinct from $1/f$ and shot noise, has been measured in systems with localized electron states \cite{Boussaad2003,Lastapis2005} and charge traps \cite{Kim2010,Cho2018b}, as well as bistable molecular conformations \cite{Baber2008,Wassel2003,Kuznetsov2007,Donhauser2001,Auwarter2011,Cho2018a,Artes2014}, and from the forming and breaking of metal-molecule bonds \cite{Nichols2010}. Using a quantum point contact  as a charge detector, Fricke et al. have also measured bimodal counting statistics, arising from telegraphic switching, in a quantum dot \cite{Fricke2007}. For molecules interacting with a vibrational mode there have been reports \cite{Koch2005} of a similar phenomenon: avalanche tunneling. Here, long periods of zero current are interrupted by phonon-assisted electron tunnelings. Lau et al. have even reported experimental measurements of avalanche tunneling in a  single-molecule graphene-fullerene transistor, and successfully modeled their results using a two-state stochastic process \cite{Lau2016}. This system, in particular, is indicative of an important transport scenario we consider: telegraphic switching arising from molecular vibrations. We show that temporal correlations arise in this scenario, and are thus potentially crucial for understanding such experimental results.

There remains, however, a dearth of theoretical literature on telegraph noise in nanoscale quantum transport. In the 1990s, Galperin et al. studied the average transparency \cite{Galperin1994} and low-frequency noise \cite{Galperin1995} through double barriers with dynamic defects. After a long gap of 20 years, theoretical quantum telegraphic switching research has resumed; Entin-Wohlman et al.\cite{Entin-Wohlman2017}, for example, used Green's functions to study quantum heat transport via a fluctuating electronic level, proposed as a model for an applied stochastic electric field. Gurvitz et al. \cite{Gurvitz2016} also used a fluctuating electronic level, but instead analyzed steady-state and transient dynamics. One of the authors (DSK) has recently investigated telegraph noise in a junction with electron-phonon interactions \cite{Kosov2018b} by adding a stochastic component to the quantum master equation.
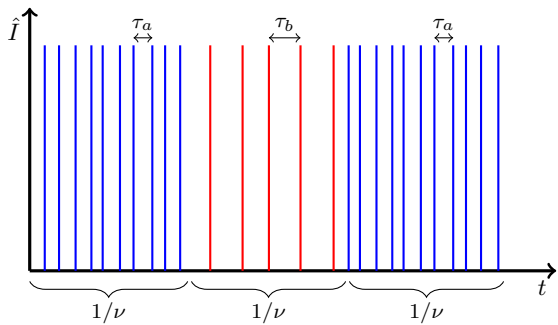
\begin{figure}
\begin{tikzpicture}
\draw[very thick,->] (0,0) -- (7,0) node[anchor=north east] {$t$};
\draw[very thick,->] (0,0) -- (0,3.5) node[anchor=north east] {$\hat{I}$};
\draw [decorate,decoration={brace,amplitude=6pt,mirror,raise=4pt},yshift=0pt]
(0,0) -- (2.1,0) node [black,midway,yshift=-16pt] {\footnotesize $1/\nu$};
\draw [decorate,decoration={brace,amplitude=6pt,mirror,raise=4pt},yshift=0pt]
(2.15,0) -- (4.2,0) node [black,midway,yshift=-16pt] {\footnotesize $1/\nu$};
\draw [decorate,decoration={brace,amplitude=6pt,mirror,raise=4pt},yshift=0pt]
(4.25,0) -- (6.3,0) node [black,midway,yshift=-16pt] {\footnotesize $1/\nu$};
\draw[thick, color = blue] (0.2,0) -- (0.2,3);
\draw[thick, color = blue] (0.39,0) -- (0.39,3);
\draw[thick, color = blue] (0.61,0) -- (0.61,3);
\draw[thick, color = blue] (0.82,0) -- (0.82,3);
\draw[thick, color = blue] (0.97,0) -- (0.97,3);
\draw[thick, color = blue] (1.2,0) -- (1.2,3);
\draw[thick, color = blue] (1.38,0) -- (1.38,3);
\draw[thick, color = blue] (1.63,0) -- (1.63,3);
\draw[thick, color = blue] (1.8,0) -- (1.8,3);
\draw[thick, color = blue] (2,0) -- (2,3);
\draw[thick, color = red] (2.4,0) -- (2.4,3);
\draw[thick, color = red] (2.83,0) -- (2.83,3);
\draw[thick, color = red] (3.18,0) -- (3.18,3);
\draw[thick, color = red] (3.6,0) -- (3.6,3);
\draw[thick, color = red] (4.04,0) -- (4.04,3);
\draw[thick, color = blue] (4.24,0) -- (4.24,3);
\draw[thick, color = blue] (4.39,0) -- (4.39,3);
\draw[thick, color = blue] (4.61,0) -- (4.61,3);
\draw[thick, color = blue] (4.82,0) -- (4.82,3);
\draw[thick, color = blue] (4.97,0) -- (4.97,3);
\draw[thick, color = blue] (5.2,0) -- (5.2,3);
\draw[thick, color = blue] (5.38,0) -- (5.38,3);
\draw[thick, color = blue] (5.63,0) -- (5.63,3);
\draw[thick, color = blue] (5.8,0) -- (5.8,3);
\draw[thick, color = blue] (6,0) -- (6,3);
\draw[thick, color = blue] (6.23,0) -- (6.23,3);
\draw[<->] (1.38,3.1) -- (1.63,3.1) node[black,midway,yshift=5pt] {\footnotesize $\tau_{a}$};
\draw[<->] (3.18,3.1) -- (3.6,3.1) node[black,midway,yshift=5pt] {\footnotesize $\tau_{b}$};
\draw[<->] (5.38,3.1) -- (5.63,3.1) node[black,midway,yshift=5pt] {\footnotesize $\tau_{a}$};
\end{tikzpicture}
\caption{Schematic of a time-series of current spikes in a system telegraphically switching, at rate $\nu$, between two characteristic first-passage times $\tau_{a}\text{ and }\tau_{b}$. Note that, to exaggerate the effects, we have reduced the stochasticity of time-intervals \textit{within} each configuration. For simplicity, each current spike represents the total current increasing by $+1$ since the last measurement.}
\label{telegraphic switching schematic}
\end{figure}

The common theme among these treatments of the telegraph noise are time-dependent stochastic additions, $\zeta(t)$, whether they be to the electronic level, $\varepsilon + U\zeta(t)$, or to the master equation itself: $\dot{\mathbf{P}}(t) = \mathbf{L}\mathbf{P}(t) + \zeta(t)\mathbf{A}\mathbf{P}(t)$. Instead, we use a general Markovian master equation with two distinct sets of states, associated with transport scenario $a$ and transport scenario $b$, connected only by a constant switching rate $\nu$. All rates are time-independent and the master equation can be solved via normal methods.

This picture of telegraph noise is particularly apt for quantum transport, as the last 20 years has produced a pantheon of fluctuation statistics that are easily calculable from the Markovian master equation. We will analyze two of these fluctuation statistics, the full counting statistics (FCS) and first-passage time distribution (FPTD), in the context of renewal theory, which is based on the renewal assumption: that successive electron tunnelings are uncorrelated and the transport is ``renewed" after each tunneling event \cite{VanKampen1981}. We expect, however, that telegraphic switching will produce strongly positive temporal correlations between successive tunnelings. The non-renewal behavior is also expected to destroy any relationships between relevant cumulants of the FCS, a fixed-time statistic, and the FPTD, a fluctuating-time statistic \cite{Ptaszynski2018}.

The FCS, for example, generates cumulants of the current distribution $\langle\langle I(t)^{k} \rangle\rangle$ over a fixed time interval $[0,t]$. The method arose \cite{Levitov1996} from the need to go beyond the average current $\langle I \rangle$ and noise $\mathcal{S}(\omega)$ to analyze fluctuations in terms of higher-order cumulants, and has been remarkably successful: describing Coulomb blockaded quantum dots \cite{Bagrets2003}, non-Markovian transport through a dissipative double quantum dot\cite{Flindt2008,Flindt2010}, and systems with electron-phonon interactions \cite{Agarwalla2015,Ueda2017}. Alongside these theoretical calculations, experimental groups have measured the FCS \cite{Gustavsson2008,Maisi2014,Gershon2008}, even up to the $15^{\text{th}}$ cumulant \cite{Flindt2009}.

The real-time single electron detection techniques required to measure the FCS have also given us access to a complementary set of statistics, which include the waiting time distribution (WTD) and the FPTD. Although the WTD has been used extensively \cite{Albert2014,Albert2011,Rudge2016a,Rudge2016b,Rudge2018,Jenei2019,Potanina2017,Tang2018,Ptaszynski2017a,Dasenbrook2015,Kosov2017b,Walldorf2018,Mi2018,Kosov2018a,Kosov2017a} in conjunction with FCS since Brandes introduced it to nanoscale transport \cite{Brandes2008}, we will use the FPTD only. The FPTD $F(n|\tau)$ is the conditional probability density that, given an electron has tunneled to the drain, the jump number first reaches $n$ after a time-delay $\tau$. Since the jump number is the $\text{\it{total}}$ number of forward and backward transitions, it directly relates to bidirectional current, whereas the WTD only works for unidirectional transport \cite{Ptaszynski2018,Rudge2019a,Rudge2019b,Singh2019}. 

In this paper, then, we consider three quantum transport telegraphic processes and search for non-renewal behavior and time correlations arising from each. In the first, we stochastically switch a magnetic field $B$ on and off an Anderson impurity, so that the electronic energy level switches between being spin-split and degenerate. Next, we model two different molecular conformations via coupling of an electronic level to two different vibrational modes. Finally, we mimic a contact forming and breaking the molecule-electrode bonds at random points in time. Previous fluctuation research using Markovian rate equations has struggled to find significant correlations between successive electron tunnelings \cite{Ptaszynski2018,Rudge2018,Rudge2019a}, since the T-matrix approach \cite{Timm2008} neglects quantum coherent effects and the Markovian baths are memory-less. In contrast, we find that, with the inclusion of telegraphic switching in the dynamics, there are significant correlations present in all scenarios.

In Section(\ref{Model Section}), we first briefly outline the general Markovian master equation and then discuss in depth all models used in our analysis. Section(\ref{Fluctuation statistics Section}) introduces the FCS and FPTD, as well as a discussion on renewal and non-renewal behavior. We present results for all three transport scenarios in Section(\ref{Results}), as well as an explanation for each, with the conclusions contained in Section(\ref{Conclusions}). 

Throughout the paper we use natural units: $\hbar = e = k_{B} = 1$.

\section{Model} \label{Model Section}

Our general model is a quantum system weakly coupled to two macroscopic electron baths. If correlations in the baths decay rapidly, then the reduced density matrix of the quantum system, $\mathbf{P}(t)$, satisfies the Markovian master equation:

\begin{align}
\dot{\mathbf{P}}(t) & = \mathbf{L}\mathbf{P}(t). \label{standard ME}
\end{align}
Here, we have mapped the $m\times m$ density matrix to an $m^{2}$ vector of which the first $m$ elements are pure states and the last $m(m-1)$ elements are coherences. The superoperator $\mathbf{L}$, the Liouvillian, thus contains all time-independent system dynamics. Off-diagonals $[\mathbf{L}]_{lk} = \Gamma_{lk}$ are the transition rate from state $k$ to state $l$, while diagonals are $[\mathbf{L}]_{kk} = -\sum\limits_{l\neq k}\Gamma_{lk}$. All transition rates are calculated via the T-matrix approach and under the secular approximation, which negelects coherences. This necessarily restricts Eq.\eqref{standard ME} to a rate equation in the pure states only and, while potentially limiting the quantum physics possible, still leaves us with a non-trivial transport regime. 

We assume that Eq.\eqref{standard ME} has a unique stationary solution: the vector $\bar{\mathbf{P}}$, which satisfies $\mathbf{L}\bar{\mathbf{P}} = 0$. In the stationary state, then, the solution of Eq.\eqref{standard ME} is

\begin{align}
\mathbf{P}(t) = e^{\mathbf{L}t}\bar{\mathbf{P}}. \label{General master equation solution}
\end{align}

We will, in fact, use the $n$-resolved probability vector $\mathbf{P}(n,t)$, whose elements $\left[\mathbf{P}(n,t)\right]_{l}$ are the probability for the system to be in state $l$ at time $t$, and for there to have been $n$ extra electrons collected in the drain in the interval $\left[0,t\right]$. It satisfies the $n$-resolved master equation:

\begin{align}
\dot{\mathbf{P}}(n,t) & = \sum_{n'}\mathbf{L}(n-n')\mathbf{P}(n,t), \label{n-resolved ME} \\
& = \mathbf{L}_{0}\mathbf{P}(n,t) + \mathbf{J}_{F}\mathbf{P}(n-1,t) + \mathbf{J}_{B}\mathbf{P}(n+1,t), \label{n-resolved ME expanded}
\end{align}
where $\mathbf{J}_{F}$ and $\mathbf{J}_{B}$ are quantum jump operators that move particles forward to the drain and backward from the drain, respectively, and $\mathbf{L}_{0} = \mathbf{L} - \mathbf{J}_{F} - \mathbf{J}_{B}$. The Fourier transform of the $n$-resolved probability vector,

\begin{align}
\mathbf{P}(\chi,t) & = \sum_{n}e^{in\chi}\mathbf{P}(n,t), \label{Fourier transformed probability vector}
\end{align}
transforms Eq.\eqref{n-resolved ME expanded} into one $1^{\text{st}}$-order differential matrix equation:

\begin{align}
\dot{\mathbf{P}}(\chi,t) & = \mathbf{L}(\chi)\mathbf{P}(\chi,t), \label{Chi dependent ME}
\end{align}
where $\mathbf{L}(\chi) = \mathbf{L}_{0} + \mathbf{J}_{F}e^{i\chi} + \mathbf{J}_{B}e^{-i\chi}$. Similarly to Eq.\eqref{General master equation solution}, the Fourier transformed $n$-resolved master equation has the solution

\begin{align}
\mathbf{P}(\chi,t) & = e^{\mathbf{L}(\chi)t}\bar{\mathbf{P}}, \label{Chi dependent ME solution}
\end{align}
where the initial condition remains the same as all measurements are performed in the stationary state.

\subsection{General telegraphic process}

Since we switch between two transport scenarios, we unfortunately cannot write a single Hamiltonian to describe the system dynamics. Rather, we will write the Hamiltonian for each scenario and then construct the master equation from ad hoc principles. The master equation for a general quantum system undergoing telegraphic switching between two scenarios $a$ and $b$ is
\begin{widetext}
\begin{align}
\frac{d}{dt}\left[\begin{array}{c} \\ \mathbf{P}_{a}(\chi,t) \\ \\ \\ \mathbf{P}_{b}(\chi,t) \\ \\ \end{array}\right] & = \left[\begin{array}{c c c c c c} 
& & & & & \\
& \mathbf{L}_{a}(\chi) - \bm{\nu} & & & \bm{\nu} & \\
& & & & & \\
& & & & & \\
& \bm{\nu} & & & \mathbf{L}_{b}(\chi) - \bm{\nu} & \\
& & & & & \\
\end{array} \right]\left[\begin{array}{c} \\ \mathbf{P}_{a}(\chi,t) \\ \\ \\ \mathbf{P}_{b}(\chi,t) \\ \\ \end{array}\right]. \label{General TS ME}
\end{align} 
\end{widetext}

From here, we will use the notation $\varphi \in [a,b]$ when referring to both of the two different transport scenarios, and $\bar{\varphi}$ when referring to the opposite scenario. Each $\mathbf{L}_{\varphi}(\chi)$ component thus refers to the Liouvillian of the $\varphi$ scenario without telegraphic switching. The matrix $\bm{\nu}$ contains the telegraphic switching rates, which are the same for each $\varphi$. They must be subtracted from $\mathbf{L}_{\varphi}(\chi)$ to conserve probability. Finally, the vector $\mathbf{P}(\chi,t) = \left[\mathbf{P}_{a}(\chi,t), \mathbf{P}_{b}(\chi,t)\right]$ is comprised of the probability distributions for the two scenarios. The transport scenarios we consider will all follow the dynamics in Eq.\eqref{General TS ME}. The jump operators are similarly defined:

\begin{align}
\mathbf{J}_{F} = \left[\begin{array}{c c c c c c} 
& & & & & \\
& \mathbf{J}_{F}^{a} & & & 0 & \\
& & & & & \\
& & & & & \\
& 0 & & & \mathbf{J}_{F}^{b} & \\
& & & & & \\
\end{array} \right], \hspace{1cm} \mathbf{J}_{B} = \left[\begin{array}{c c c c c c} 
& & & & & \\
& \mathbf{J}_{B}^{a} & & & 0 & \\
& & & & & \\
& & & & & \\
& 0 & & & \mathbf{J}_{B}^{b} & \\
& & & & & \\
\end{array} \right]. \label{TS Jump Operators}
\end{align}

\subsection{Anderson impurity}

Our first scenario is the well-known Anderson impurity model:

\begin{align}	
H_{M} & = \sum_{\sigma\in\{\uparrow,\downarrow\}} \varepsilon_{\sigma}a^{\dagger}_{\sigma}a^{ }_{\sigma} + U n_{\uparrow}n_{\downarrow}. \label{Anderson Hamiltonian}	
\end{align}

First introduced to describe local magnetic impurities in metals\cite{Anderson1961}, Eq.\eqref{Anderson Hamiltonian} has found use in molecular electronics theory, where $U$ generally describes repulsive electron-electron interactions within the orbital. The impurity is also coupled to two macroscopic metal elctrodes, the source (S) and drain (D), with combined Hamiltonian

\begin{align}
H_{\text{electrodes}} = \sum_{\alpha = S,D}\sum_{k}\varepsilon^{}_{\alpha,k}a^{\dag}_{\alpha,k}a^{ }_{\alpha,k}, \label{Electrodes Hamiltonian}
\end{align}
and interaction Hamiltonian,
\begin{align}
H_{T} = \sum_{\alpha = S,D}\sum_{k}t^{}_{\alpha,k}(a^{\dag}_{\alpha,k}a^{}+a^{\dag}a^{}_{\alpha,k}), \label{Interaction Hamiltonian}
\end{align}
enabling tunneling of electrons between the electrode-molecule configuration. The operators $a^{\dag}_{\alpha,k}$ and $a^{}_{\alpha,k}$ create and annihilate electrons in state $k$ in electrode $\alpha$, and $t_{\alpha,k}$ is the tunneling matrix element between the molecular orbital and state $k$ in electrode $\alpha$. Combined, the Hamiltonian of the entire system is 

\begin{equation}
H= H_{M} + H_{\text{electrodes}} + H_{T}. \label{Total Hamiltonian}
\end{equation}

We will use the Anderson impurity to model switching between scenarios $a$ and $b$: a molecular orbital without an applied magnetic field and a molecular orbital with an applied magnetic field $B$, respectively. In the absence of a magnetic field, and barring further fine splitting, spin-$\uparrow$ and spin-$\downarrow$ electrons require the same charging energy to enter the molecule: $\varepsilon^{a}_{\uparrow} = \varepsilon^{a}_{\downarrow} = \varepsilon_{0}$. Once the magnetic field is applied, however, the spin-split energies are $\varepsilon^{b}_{\downarrow} = \varepsilon_{0} + B/2$ and $\varepsilon^{b}_{\uparrow} = \varepsilon_{0} - B/2$.

We combine these two scenarios in the probability vector 
\begin{widetext}
\begin{align}
\mathbf{P}(\chi,t) = [P^{a}_{0}(\chi,t),P^{a}_{\uparrow}(\chi,t),P^{a}_{\downarrow}(\chi,t),P^{a}_{2}(\chi,t),P^{b}_{0}(\chi,t),P^{b}_{\uparrow}(\chi,t),P^{b}_{\downarrow}(\chi,t),P^{b}_{2}(\chi,t)]^{T}. \label{Anderson chi probability vector}
\end{align}
\end{widetext}
For sequential tunneling only, and under the Born-Markov approximation \cite{Mitra2004}, the $\chi$-dependent Liouvillian of scenario $\varphi$ is

\begin{widetext}
\begin{align}
	\mathbf{L}_{\varphi}(\chi) & = \left[\begin{array}{c c c c}
-(\Gamma^{\varphi}_{\uparrow 0}+\Gamma^{\varphi}_{\downarrow 0}) & \Gamma^{\varphi}_{0\uparrow}(\chi) & \Gamma^{\varphi}_{0\downarrow}(\chi) & 0 \\
		\\
		\Gamma^{\varphi}_{\uparrow 0}(\chi) & -(\Gamma^{\varphi}_{0\uparrow} + \Gamma^{\varphi}_{2\uparrow}) & 0 & \Gamma^{\varphi}_{\uparrow 2}(\chi) \\
		\\
		\Gamma^{\varphi}_{\downarrow 0}(\chi) & 0 & -(\Gamma^{\varphi}_{0\downarrow} + \Gamma^{\varphi}_{2\downarrow}) & \Gamma^{\varphi}_{\downarrow 2}(\chi) \\
		\\
		0 & \Gamma^{\varphi}_{2\uparrow}(\chi) & \Gamma^{\varphi}_{2\downarrow}(\chi) & -(\Gamma^{\varphi}_{\uparrow 2} + \Gamma^{\varphi}_{\downarrow 2}) \\
		\\
	\end{array}\right],
\end{align}
\end{widetext}
and the jump operators are
\begin{widetext}
\begin{align}
	\mathbf{J}^{\varphi}_{F} = \left[\begin{array}{c c c c}
		0 & \Gamma_{0\uparrow}^{D,\varphi} & \Gamma_{0\downarrow}^{D,\varphi} & 0 \\
		0 & 0 & 0 & \Gamma_{\uparrow 2}^{D,\varphi} \\
		0 & 0 & 0 & \Gamma_{\downarrow 2}^{D,\varphi} \\
		0 & 0 & 0 & 0 \\
		\end{array}\right] \hspace{0.5cm} \text{ and } \hspace{0.5cm} \mathbf{J}^{\varphi}_{B} = \left[\begin{array}{c c c c}
		0 & 0 & 0 & 0 \\
		\Gamma_{\uparrow 0}^{D,\varphi} & 0 & 0 & 0 \\
		\Gamma_{\downarrow 0}^{D,\varphi} & 0 & 0 & 0 \\
		0 & \Gamma_{2\uparrow}^{D,\varphi} & \Gamma_{2\downarrow}^{D,\varphi} & 0 \\
		\end{array}\right].
\end{align}
\end{widetext}

The total $\chi$-dependent rates contain a source and drain component:

\begin{align}
\Gamma^{\varphi}_{lk} & = \Gamma_{lk}^{S,\varphi} + \Gamma_{lk}^{D,\varphi}e^{\pm i\chi},
\end{align}
where the $\pm$ is positive if the fermionic occupation increases from state $k$ to state $l$, and negative if the fermionic occupation decreases. Explicitly now, the spin-dependent rates are 
\begin{align}
\Gamma_{\sigma0}^{\alpha,\varphi} & = \gamma^{\alpha,\varphi} \: n_{F}(\varepsilon^{\varphi}_{\sigma}-\mu_{\alpha}), \\
\Gamma_{0\sigma}^{\alpha,\varphi} & = \gamma^{\alpha,\varphi} \: \big{(}1-n_{F}(\varepsilon^{\varphi}_{\sigma}-\mu_{\alpha})\big{)}, \\
\Gamma_{\sigma2}^{\alpha,\varphi} & = \gamma^{\alpha,\varphi} \: \big{(}1-n_{F}(\varepsilon^{\varphi}_{\sigma}+U-\mu_{\alpha})\big{)}, & \text{and} \\
\Gamma_{2\sigma}^{\alpha,\varphi} & = \gamma^{\alpha,\varphi} \: n_{F}(\varepsilon^{\varphi}_{\sigma}+U-\mu_{\alpha}). \\
\nonumber\end{align}

\subsection{Holstein model}

We also consider a molecule changing conformations, and thus changing its vibrational interactions. Conformation $\varphi$ is assumed to be a single molecular orbital $\varepsilon^{\varphi}_{0}$ interacting with a vibrational mode $\omega_{\varphi}$.  This is the well-known Holstein model:

\begin{equation}
H_{M}= \varepsilon_0 a_{}^{\dag} a  + \lambda \omega (b_{}^{\dag} + b) a_{}^{\dag} a +  \omega b^{\dag}_{} b:
\end{equation}
where $a_{}^{\dag}$ and $a$ are the fermionic creation and annihilation operator, respectively; $b_{}^{\dag}$ and $b$ are the bosonic creation and annihilation operators, respectively; and $\lambda$ denotes the electron-phonon coupling. We ignore spin-degeneracy by working in the Coulomb blockade regime. We note that the molecule is still coupled to two electrodes, with Hamiltonians in Eq.\eqref{Electrodes Hamiltonian}, interaction Hamiltonians in Eq.\eqref{Interaction Hamiltonian}, and total system Hamiltonian in Eq.\eqref{Total Hamiltonian}.

We apply the canonical Lang-Firsov transformation \cite{Lang1963} to diagonalize the molecular Hamiltonian:
\begin{align}
H_{M} & = \varepsilon \tilde{a}^{\dag}\tilde{a} + \omega \tilde{b}^{\dag}\tilde{b}, \label{LF Hamiltonian}
\end{align}
which renormalizes the orbital energy to $\varepsilon = \varepsilon_{0} - \frac{\lambda^{2}}{\omega}$. The eigenstates of Eq.\eqref{LF Hamiltonian}, $|mq\rangle$, denote occupation of $m = \left\{0,1\right\}$ electrons and $q = \left\{0,1,2, \hdots , +\infty\right\}$ phonons, with associated eigenenergies $E_{mq} = \varepsilon m + \omega q$. 

Since the system switches between two configurations, there will be two sets of parameters: $\{\varepsilon_{0}^{a},\lambda_{a},\omega_{a}\}$ and $\{\varepsilon_{0}^{b},\lambda_{b},\omega_{b}\}$. We therefore seek a master equation for the probability that at time $t$ the system is occupied by $m$ electrons and $q$ phonons, while in configuration $\varphi$, which is denoted $P_{mq}^{\varphi}(t)$:

\begin{widetext}
\begin{align}
\dot{P}^{\varphi}_{0q}(t) & = \nu_{0}\left(P^{\bar{\varphi}}_{0q}(t) - P^{\varphi}_{0q}(t)\right) + \sum_{\alpha q'} \Gamma^{\alpha,\varphi}_{0q,1q'} P^{\varphi}_{1q'} (t)  -  \Gamma^{\alpha,\varphi}_{1q', 0q} P^{\varphi}_{0q}(t),
\label{me1_a}
\\
\dot{P}^{\varphi}_{1q}(t) & = \nu_{1}\left(P^{\bar{\varphi}}_{1q}(t) - P^{\varphi}_{1q}(t)\right) + \sum_{\alpha q'} \Gamma^{\alpha,\varphi}_{1q,0q'} P^{\varphi}_{0q'}(t)  -  \Gamma^{\alpha,\varphi}_{0q',1q} P^{\varphi}_{1q}(t),
\label{me2_a}
\end{align}
\end{widetext}
The molecule switches between the two vibrational modes with rate $\nu_{0}$, when the system is electronically empty, and rate $\nu_{1}$, when the system is singly occupied. Transitions within the configurations obey the usual rules;

\begin{align}
\Gamma^{\alpha,\varphi}_{0q',1q} & =  \gamma^{\alpha,\varphi} |X^{\varphi}_{q'q}|^2 \left[1-n_{F}(\varepsilon_{\varphi}-\omega_{\varphi} (q'-q) - \mu_{\alpha}) \right] \label{TR 01}
\end{align}
is the transition rate from state $|1q\rangle_{\varphi}$ to state $|0q'\rangle_{\varphi}$, via tunneling to electrode $\alpha$, and likewise 
\begin{align} 
\Gamma^{\alpha,\varphi}_{1q',0q} & =  \gamma^{\alpha,\varphi} |X^{\varphi}_{q'q}|^2  n_{F}\left(\varepsilon_{\varphi}+\omega_{\varphi} (q'-q)-\mu_\alpha\right). \label{TR 10}
\end{align}
is the transition rate between $|0q\rangle_{\varphi}$ and $|1q'\rangle_{\varphi}$. The energy level $\varepsilon_{\varphi}$ is broadened by the factor $\gamma^{\alpha} = 2\pi|t_{\alpha}|^{2}\rho(\varepsilon_{\varphi})$, where the density of states $\rho(\varepsilon_{\varphi})$ is assumed to be constant. In Eq.\eqref{TR 01} and Eq.\eqref{TR 10} the transition rates also depend on the Fermi-Dirac occupation 
\begin{align}
n_{F}(E-\mu_{\alpha}) & = \frac{1}{1+e^{(E-\mu_\alpha)/T}},
\end{align}
the electrode temperature $T$, and the $\alpha$-electrode chemical potential $\mu_{\alpha}$. Transitions between different phonon states are determined by the Franck-Condon factor $X^{\varphi}_{qq'}$:
\begin{align}
X^{\varphi}_{qq'} & = \langle q | e^{-\lambda_{\varphi}(b^{\dag}-b)}|q'\rangle.
\end{align}

The Fourier transformed $n$-resolved probability vector is 
\begin{widetext}
\begin{align}
\mathbf{P}(\chi,t) & = \left[P^{a}_{00}(\chi,t), P^{a}_{10}(\chi,t), \hdots, P^{a}_{0N}(\chi,t), P^{a}_{1N}(\chi,t), P^{b}_{00}(\chi,t), P^{b}_{10}(\chi,t), \hdots, P^{b}_{0N}(\chi,t), P^{b}_{1N}(\chi,t)\right],
\end{align}
\end{widetext}
where $N$ is the maximum number of phonons included in the transport, chosen such that $N\omega_{\varphi} \gg V_{SD},\gamma,T$. $\mathbf{P}(\chi,t)$ therefore has length $4(N+1)$ and its components $\mathbf{P}^{\varphi}(\chi,t)$ follow the master equation:

\begin{widetext}
\begin{align}
\dot{P}^{\varphi}_{0q}(\chi,t) & = \nu_{0}\left(P^{\bar{\varphi}}_{0q}(\chi,t) - P^{\varphi}_{0q}(\chi,t)\right) + \sum_{q'} \left(\Gamma^{S,\varphi}_{0q;1q'} + \Gamma^{D,\varphi}_{0q;1q'}e^{i\chi}\right)P^{\varphi}_{1q'}(\chi,t) - \sum_{\alpha q'} \Gamma^{\alpha,\varphi}_{1q';0q}P^{\varphi}_{0q}(\chi,t), \label{P0 rate equation holstein a} \\ 
\dot{P}^{\varphi}_{1q}(\chi,t) & = \nu_{1}\left(P^{\bar{\varphi}}_{1q}(\chi,t) - P^{\varphi}_{1q}(\chi,t)\right) + \sum_{q'} \left(\Gamma^{S,\varphi}_{1q;0q'} + \Gamma^{D,\varphi}_{1q;0q'}e^{-i\chi}\right)P^{\varphi}_{0q'}(\chi,t) - \sum_{\alpha q'} \Gamma^{\alpha,\varphi}_{0q';1q}P^{\varphi}_{1q}(\chi,t). \label{P1 rate equation holstein a}
\end{align}
\end{widetext}
From here the $\chi$-dependent master equation can easily be split into the quantum jump operators, which are constructed according to Eq.\eqref{TS Jump Operators}. The individual $\mathbf{J}_{F}^{\varphi}$ and $\mathbf{J}_{B}^{\varphi}$ are also easily defined, as in Ref.[\onlinecite{Rudge2019a}]. At this point the full Liouvillian remains too large to be written in matrix form, since we have made no assumptions about the underlying phonon distribution. If the phonons are in thermal equilibrium with an external bath at temperature $T_{V}$, however, then they must be Boltzmann distributed, and $P^{\varphi}_{nq}(\chi,t)$ can be factorized:
\begin{align}
P^{\varphi}_{nq}(\chi,t) & = P^{\varphi}_{n}(\chi,t)\frac{e^{-q\omega_{\varphi}/T_{V}}}{1-e^{-\omega_{\varphi}/T_{V}}}. \label{ansatz}
\end{align}
For equilibrated phononons, therefore, we define effective transition rates using the ansatz in Eq.\eqref{ansatz},
\begin{align}
T^{\varphi}_{lk} & = \sum_{\alpha} T^{\alpha,\varphi}_{lk} \\
& = \sum_{\alpha,qq'} \Gamma^{\alpha,\varphi}_{lq;kq'} \frac{e^{-q\omega_{\varphi}/T_{V}}}{1-e^{-\omega_{\varphi}/T_{V}}},
\end{align}
which define the corresponding master equation,
\begin{widetext}
\begin{align}
\mathbf{L}(\chi) & = \left[\begin{array}{cccc} -(T^{a}_{10}+\nu_{0}) & T_{01}^{S,a} + T_{01}^{D,a}e^{i\chi} & \nu_{0} & 0 \\
\\ 
T^{S,a}_{10} + T^{D,a}_{10}e^{-i\chi} & -(T_{01} + \nu_{1}) & 0 & \nu_{1} \\
\\
\nu_{0} & 0 & -(T^{b}_{10}+\nu_{0}) & T_{01}^{S,b} + T_{01}^{D,b}e^{i\chi} \\ 
\\
0 & \nu_{1} & T^{S,b}_{10} + T^{D,b}_{10}e^{-i\chi} & -(T^{b}_{01} + \nu_{1}) \end{array}\right], \label{equilibrated Lo}
\end{align}
\end{widetext}
and jump operators:
\begin{align}
\mathbf{J}_{F} = \left[\begin{array}{cccc} 0 & T_{01}^{D,a} & 0 & 0 \\ 
										  0 & 0 & 0 & 0 \\
										  0 & 0 & 0 & T_{01}^{D,b} \\ 
										  0 & 0 & 0 & 0 \\ \end{array}\right] \hspace{0.15cm} \text{and} \hspace{0.15cm}
\mathbf{J}_{B}  = \left[\begin{array}{cccc} 0 & 0 & 0 & 0 \\ 
										  T_{10}^{D,a} & 0 & 0 & 0 \\
										  0 & 0 & 0 & 0 \\ 
										  0 & 0 & T_{10}^{D,b} & 0 \\ \end{array}\right].
\end{align}

\section{Fluctuation statistics} \label{Fluctuation statistics Section}

\subsection{FCS}

We start this section with the FCS, which we calculate as cumulants of the distribution of transferred charge: $P(n,t) = \left(\mathbf{I},\mathbf{P}(n,t)\right)$, where $\mathbf{I}$ is a row vector of ones the same length as $\mathbf{P}(n,t)$. The moment generating function (MGF) of $P(n,t)$ is

\begin{align}
M(\chi,t) & = \sum_{n=0}^{\infty} e^{in\chi} P(n,t) \label{Chi inner product} \\
& = \left(\mathbf{I},\mathbf{P}(\chi,t)\right),
\end{align}
where the second line follows by comparing with the Fourier transform in Eq.\eqref{Fourier transformed probability vector}. We will, in fact, seek the cumulant generating function (CGF) $K(\chi,t) = \ln M(\chi,t)$, from which successive cumulants $\langle\langle I^{k}\rangle\rangle$ can be calculated, after inserting the solution from Eq.\eqref{Chi dependent ME solution}:
\begin{align}
\langle\langle I(t)^{k}\rangle\rangle & = \frac{d}{dt} (-i)^{k} \frac{\partial^{k}}{\partial\chi^{k}}\ln\left(\mathbf{I},e^{\mathbf{L}(\chi)t}\bar{\mathbf{P}}\right)\Bigg|_{\chi=0}. \label{Full time FCS}
\end{align}
In the long-time, or large-deviation, limit, the CGF is dominated by the eigenvalue of $\mathbf{L}(\chi)$ with the largest real part \cite{Bagrets2003,Nazarov2003}: $\lim\limits_{t\rightarrow\infty}K(\chi,t) = t\Lambda_{max}(\chi)$. The current cumulants, then, are the time-independent asymptotic rates of the cumulants of $P(n,t)$:

\begin{align}
\langle\langle I^{k}\rangle\rangle & = (-i)^{k} \frac{\partial^{k}}{\partial\chi^{k}} \lambda_{max}(\chi)\Bigg|_{\chi=0}. \label{FCS generating function}
\end{align}
The long-time limit is generally used, as the full expression in Eq.\eqref{Full time FCS} is difficult to evaluate for most systems. 

The first cumulant $\langle\langle I \rangle\rangle$ is just the stationary current, which, although useful, does not provide information on fluctuations. The famous Fano factor, defined from the zero-frequency noise $\mathcal{S}(0)$,

\begin{align}
F & = \frac{\mathcal{S}(0)}{2\langle I \rangle} \\
& = \frac{\langle\langle I ^{2}\rangle\rangle}{\langle I \rangle}, \label{FF n definition}
\end{align}
provides information on the relative $\text{\it{width}}$ of the distribution. The Fano factor scales the current variance in terms of a Poissonian distribution: if $F=1$, the transport is Poissonian; if $F<1$, the transport is sub-Poissonian; and if $F>1$, the transport is super-Poissonian.

\subsection{First-passage time distribution}

The FPTD $F(n|t_{0},t_{0}+\tau)$ is the probability density that the jump number first reaches $n$ after a time delay $\tau$, conditioned upon the initial probability that an electron tunnels to the drain when counting begins at $t_{0}$. In the steady state, the initial $t_{0}$ is arbitrary and the FPTD depends only on the time-delay: $F(n|\tau)$. For a rigorous derivation, we direct the reader to Ref.[\onlinecite{Ptaszynski2018}] and Ref.[\onlinecite{Rudge2019b}]; however, we provide a brief summary below.

The FPTD definition rests on the transition matrix $\mathbf{T}(n-n',t-t')$, which contains conditional probabilities that map some probability distribution $\mathbf{P}(n',t')$ to a distribution at a later time $\mathbf{P}(n,t)$:

\begin{align}
\mathbf{P}(n,t) & = \mathbf{T}(n-n',t-t')\mathbf{P}(n',t'). \label{transition matrix demo}
\end{align}
We can now separate this process using the FPTD. Consider the probability vector $\mathbf{P}(n,t)$, which follows Eq.\eqref{transition matrix demo} when $n'=t'=0$:
\begin{align}
\mathbf{P}(n,t) & = \mathbf{T}(n,t)\mathbf{P}(0,0). \label{transition matrix demo}
\end{align}
 We can also obtain $\mathbf{P}(n,t)$, however, by 
\begin{align}
\mathbf{P}(n,t) & = \int_{0}^{\infty} \: d\tau \: \mathbf{T}(0|t-\tau)\mathbf{F}(n|\tau). \label{P from FPTD}
\end{align}
Here, $\mathbf{F}(n|\tau)$ is a vector of first-passage time probabilities distributed over the molecular states. We multiply the first-passage time probability by the conditional probability that, given the jump number is $n$ at time $\tau$, it does not change in the interval $\left[\tau,t\right]$. Of course, we must also integrate over all possible first-passage times $\tau$. 

Combining Eq.\eqref{transition matrix demo} and Eq.\eqref{P from FPTD}, and taking a Laplace transform, we get 
\begin{align}
\tilde{\mathbf{F}}(n|z) & = \tilde{\mathbf{T}}(0|z)\tilde{\mathbf{T}}(n,z)\mathbf{P}(0). \label{time space FPTD eqn}
\end{align}
Since molecular probabilities are normalised, summing all elements of $\tilde{\mathbf{F}}(n|z)$ must yield the Laplace transform of the FPTD:
\begin{align}
\tilde{F}(n|z) & = \left(\mathbf{I},\tilde{\mathbf{T}}(0|z)\tilde{\mathbf{T}}(n,z)\mathbf{P}(0)\right). \label{time space FPTD eqn}
\end{align}

The transition matrix $\tilde{\mathbf{T}}(n,z)$, which we evaluate numerically, originates from the dynamics contained in the Liouvillian \cite{Saito2016,Ptaszynski2018,Rudge2019a,Rudge2019b},
\begin{align}
\tilde{\mathbf{T}}(n|z) & = \frac{1}{2\pi}\int_{0}^{2\pi} d\chi e^{-in\chi} \left[z - \mathbf{L}(\chi),\right]^{-1}, \label{Transition matrix laplace definition}
\end{align}
but we must always choose the initial probability vector $\mathbf{P}(0)$ ourselves. To keep our analysis comparable to standard fluctuating statistics, for all calculations we choose
\begin{align}
\mathbf{P}(0) & = \frac{\mathbf{J}_{F}\bar{\mathbf{P}}}{\left(\mathbf{I},\mathbf{J}_{F}\bar{\mathbf{P}}\right)}.
\end{align}

Note that we do not have to define an analogous FPTD for tunnelings {\it from} the drain, as the jump number $n$ is the sum of forward and backward transitions and so is naturally bidirectional. The $k^{th}$ cumulant of $F(n|\tau)$ is easily calculated from $\tilde{F}(n|z)$:
\begin{align}
\langle\langle\tau^{k}_{n}\rangle\rangle & = (-1)^{k} \lim_{z\rightarrow 0^{+}}\left[\frac{d^{k}}{dz^{k}}\ln\tilde{F}(n|z)\right]. \label{CGF FPTD}
\end{align}
Since $\mathbf{L}(\chi)$ is singular for $\chi = \{0,2\pi\}$ \cite{Ptaszynski2018}, we need to take the limit $z\rightarrow 0^{+}$ in Eq.\eqref{CGF FPTD}. As we do with the current, we focus on the first and second cumulants. They combine to form the randomness parameter, which is analogous to the Fano factor:
\begin{align}
R_{n} & = \frac{\langle\langle\tau_{n}^{2}\rangle\rangle}{\left(\langle\tau_{n}\rangle\right)^{2}}, \label{RP FPTD definition}
\end{align}

\subsection{Renewal and non-renewal theory}

Recent work in nanoscale fluctuation statistics has spurred an interest in renewal theory, which examines the relationships between fixed-time and fluctuating-time statistics. If the renewal assumption is satisfied, then subsequent first-passage times are uncorrelated and the joint FPTD factorises \citep{Ptaszynski2018,Rudge2019b}:
\begin{align}
F(n|\tau_{n} ; n'|\tau_{n'}) = F(n|\tau_{n}) \: F(n' -  n|\tau_{n'} - \tau_{n}), \label{FPTD renewal 1}
\end{align}
or equivalently
\begin{align}
F(n|\tau) = F(1|\tau)^{n} \label{FPTD renewal 2}.
\end{align}
Here, we have introduced the second-order FPTD $F(n|\tau_{n} ; n'|\tau_{n'})$: the probability density that the jump number first reaches $n$ after a time $\tau$, and first reaches $n'$ after a time $\tau_{n'}$, conditioned upon an original jump to the drain. Since we consider only sequential transitions, if $n' > n$ then $\tau' > \tau$. The renewal assumption in Eq.\eqref{FPTD renewal 2}, combined with the cumulant definition in Eq.\eqref{CGF FPTD}, produces a linear relation between FPTD cumulants:
\begin{align}
\langle\langle\tau^{k}_{n}\rangle\rangle & = (-1)^{k} \lim_{z\rightarrow 0^{+}}\left[\frac{d^{k}}{dz^{k}}\ln\tilde{F}(1|z)^{n}\right], \\
& = n \langle\langle\tau^{k}_{1}\rangle\rangle. \label{FPTD cumulant relations}
\end{align}

The simplification in Eq.\eqref{FPTD cumulant relations} is the key to connecting fixed-time and fluctuating-time statistics. It is well-established in quantum transport theory that, if the renewal assumption is satisfied, exact relations exist between the FCS and equivalent cumulants of fluctuating-time distributions \cite{Brandes2008,Budini2011,Albert2011,Rudge2019b}, including the FPTD \cite{Ptaszynski2018,Rudge2019b}.

In the long-time limit, for example, the average current is constructed from the average first-passage time as
\begin{align}
\langle I \rangle & = \lim_{n\rightarrow\infty} \frac{n}{\langle\tau_{n}\rangle}.
\end{align}
But the average first passage time for $n$ electrons can be expressed using the average first-passage time of a single electron 
\begin{equation}
 \langle\tau_{n}\rangle = n\langle \tau_{1}\rangle
\end{equation}
only if the renewal relation in Eq.\eqref{FPTD cumulant relations} is satisfied or electron transport is unidirectional \cite{Rudge2019b}; the current becomes
\begin{align}
\langle I \rangle & =  \frac{1}{\langle\tau_{1}\rangle}.
\end{align}
Similarly, if the renewal assumption is satisfied, the randomness parameter and Fano factor are also equal:
\begin{align}
\frac{\langle\langle I^{2} \rangle\rangle}{\langle I \rangle} & = n \frac{\langle\langle \tau_{n}^{2} \rangle\rangle}{\langle \tau_{n} \rangle^{2}} = \frac{\langle\langle \tau_{1}^{2} \rangle\rangle}{\langle \tau_{1} \rangle^{2}}.
\end{align}
So we see that the fluctuating- and fixed-time statistics provide a direct test of the renewal assumption; if we plot the FCS and FPTD cumulants alongside one another and find where they coincide, we will have found a regime of renewal transport. In the opposite case, when the FCS are not reproduced by the FPTD cumulants, then we will have found a regime of non-renewal transport.
 
In this regime, we would by definition expect that there are correlations between successive first-passage times: information unavailable from the zero-frequency counting statistics. These are quantified by the Pearson correlation coefficient, defined here for successive first-passage times $\tau$ and $\tau'$:
\begin{align}
p  & = \frac{\langle\tau\tau'\rangle - \langle\tau\rangle^{2}}{\langle\langle\tau^{2}\rangle\rangle}. \label{Correlation coefficient}
\end{align}
We see that if the renewal assumption is satisfied then $\langle\tau\tau'\rangle = \langle\tau\rangle^{2}$ and the Pearson coefficient is formally zero. Unfortunately, the Pearson coefficient $p$ is not easily defined from the second-order FPTD $F(n'|\tau';n|\tau)$. Instead, Ptaszynski \cite{Ptaszynski2018} has developed a method from $F(2|\tau)$:
\begin{align}
p & = \frac{\langle\langle \tau_{2}^{2} \rangle\rangle}{2\langle\langle \tau_{1}^{2} \rangle\rangle} - 1. \label{Pearson FPTD}
\end{align}
In Eq.\eqref{Pearson FPTD} the correlation is between $\tau_{1}$, when $n=1$ for the first time, and $\tau_{1'} =\tau_{2} - \tau_{1}$, the time $\text{\it{delay}}$ until $n=2$ for the first time. 

\section{Results} \label{Results}

We will start this section with a discussion on the fluctuation behavior expected from telegraph noise. Regardless of the underpinning Hamiltonians, all scenarios we analyze follow a simple premise. The molecular system $H_{m}$ randomly switches between two configurations with two distinct sets of transport parameters. We expect, therefore, that each transport configuration has an associated characteristic current $\langle I \rangle_{a}\text{ and }\langle I \rangle_{b}$ and an associated characteristic first-passage time $\langle \tau \rangle_{a}\text{ and }\langle \tau \rangle_{b}$. 

If the transport parameters are set such that these characteristic first-passage times are appreciably different, and the switching rate between configurations is small enough that the transport tends to get ``stuck" in each for a long amount of time, then the dynamics will be quantitatively similar to Fig.(\ref{telegraphic switching schematic}). In it, there are relatively long periods where the first-passage times are clustered around $\langle\tau\rangle_{a}$ and then relatively long periods where the first-passage times are clustered around $\langle\tau\rangle_{b}$. That is, if a first-passage time close to $\langle\tau\rangle_{a}$ is recorded, then the next first-passage time is likely to be close to $\langle\tau\rangle_{a}$ as well, and likewise for $\langle\tau\rangle_{b}$. The transport should thus be accompanied by positive correlations between successive first-passage times.

We note that all results comparing the Fano factor and the randomness parameter refer to $R = R_{1}$, the randomness parameter of $F(1|\tau)$.

\subsection{Magnetic switching: Anderson model}

\begin{figure*}
		\begin{subfigure}{0.4\textwidth}
			\centering
			\includegraphics[width=\textwidth]{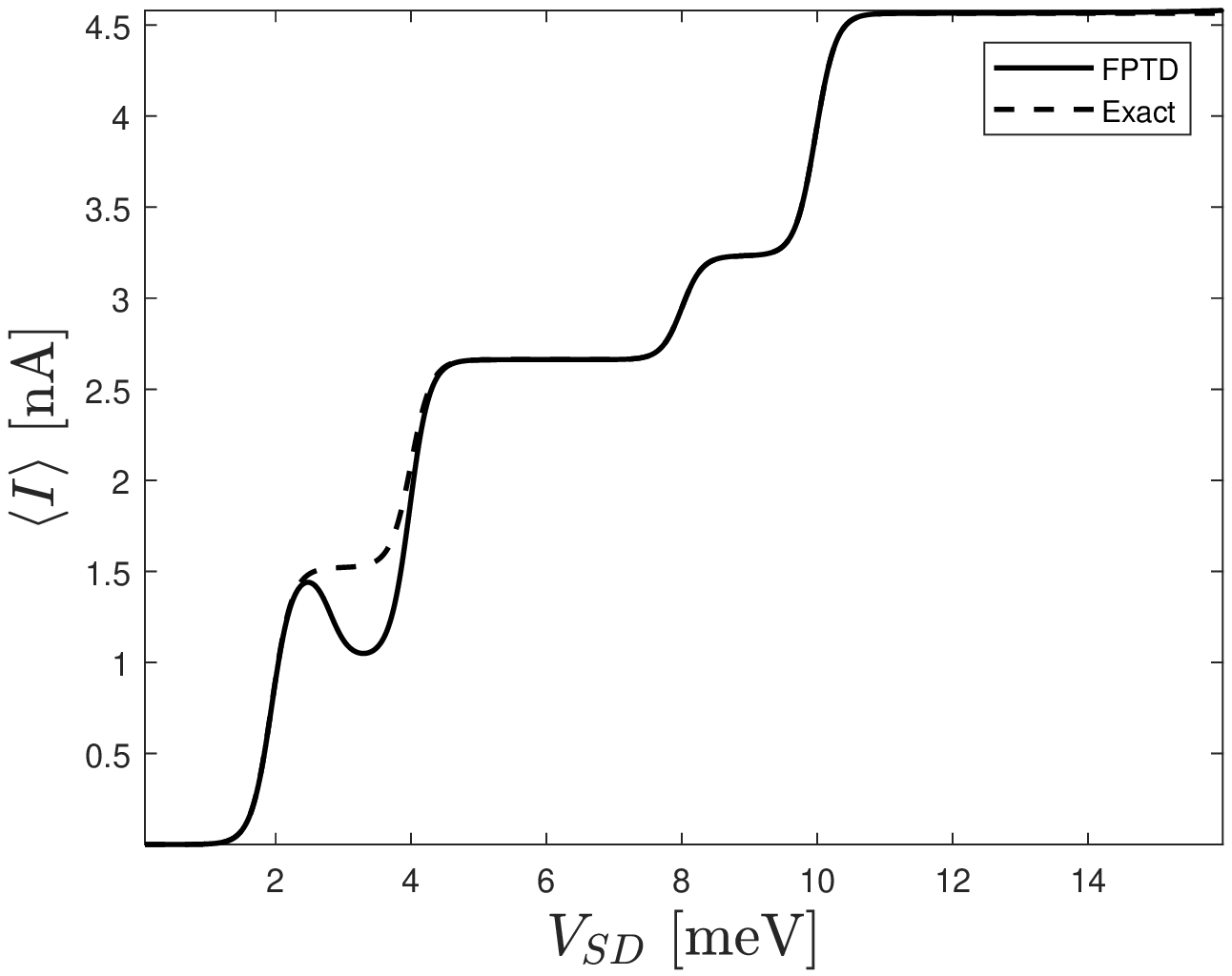}
			\caption{} 
			\label{Current Anderson U = 1}
		\end{subfigure} \: \begin{subfigure}{0.4\textwidth}
			\centering
			\includegraphics[width=\textwidth]{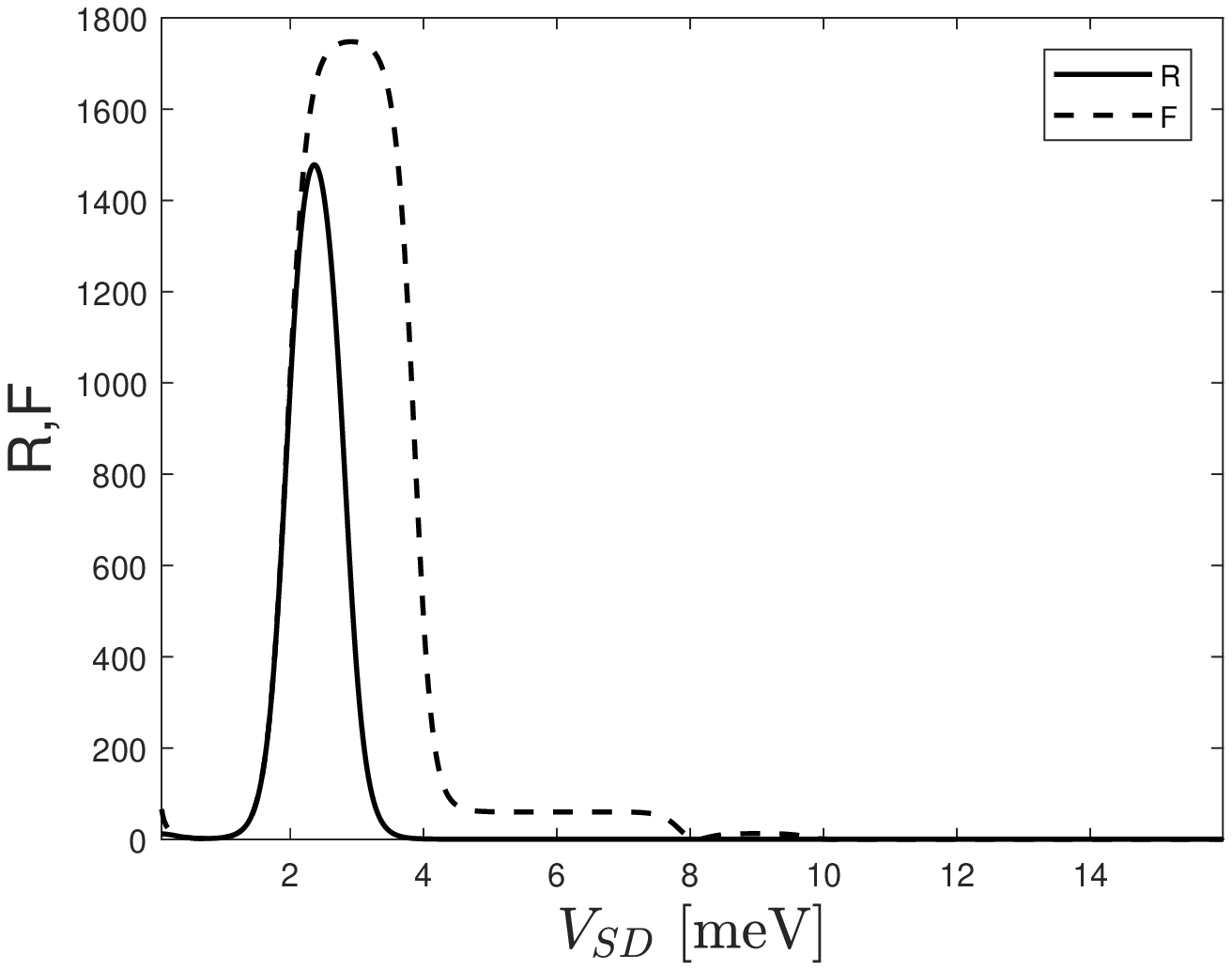} 
			\caption{}
			\label{Fano Anderson U = 1}
		\end{subfigure}
    \caption{Comparison of the first (a) and second (b) cumulants of the FCS and the FPTD as a function of $V_{SD}$. The spin-degenerate energy level is $\varepsilon_{0} = 1\text{meV}$, the magnetic field is $B/2 = 3\text{meV}$, the Coulomb repulsion is $U = 1\text{meV}$, the S-D temperature is $T = 75\mu\text{eV}$, and $\gamma^{\alpha,\varphi} = \frac{\gamma}{2}$. All telegraphic switching rates are equal: $\nu_{k} = \nu = 10^{-4}\gamma$, where $k \in \{0,\uparrow,\downarrow,2\}$.}
\end{figure*}

The current plot in Fig.(\ref{Current Anderson U = 1}) displays the $I-V$ characteristics we expect from an Anderson impurity undergoing telegraphic switching. The current undergoes steplike increases as $V_{SD}$ approaches each energy level: $\varepsilon_{0},\varepsilon_{0}\pm  B/2,\text{ and }\varepsilon_{T}$. The step at $V_{SD}/2 = 1\text{meV}$ is larger as it corresponds to the $\varepsilon^{a}_{\uparrow}\text{ and }\varepsilon^{a}_{\downarrow}$ levels simultaneously opening, while the step at $V_{SD}/2 = 5\text{meV}$ corresponds to the double level $\varepsilon_{T}$ opening for both scenarios. The FPTD largely mimics this behavior except for a region between $1\text{meV} \leq V_{SD}/2 \leq 2\text{meV}$: a regime of non-renewal behavior.

\begin{figure*}
    \begin{subfigure}{0.49\textwidth}
	\includegraphics[width=\textwidth]{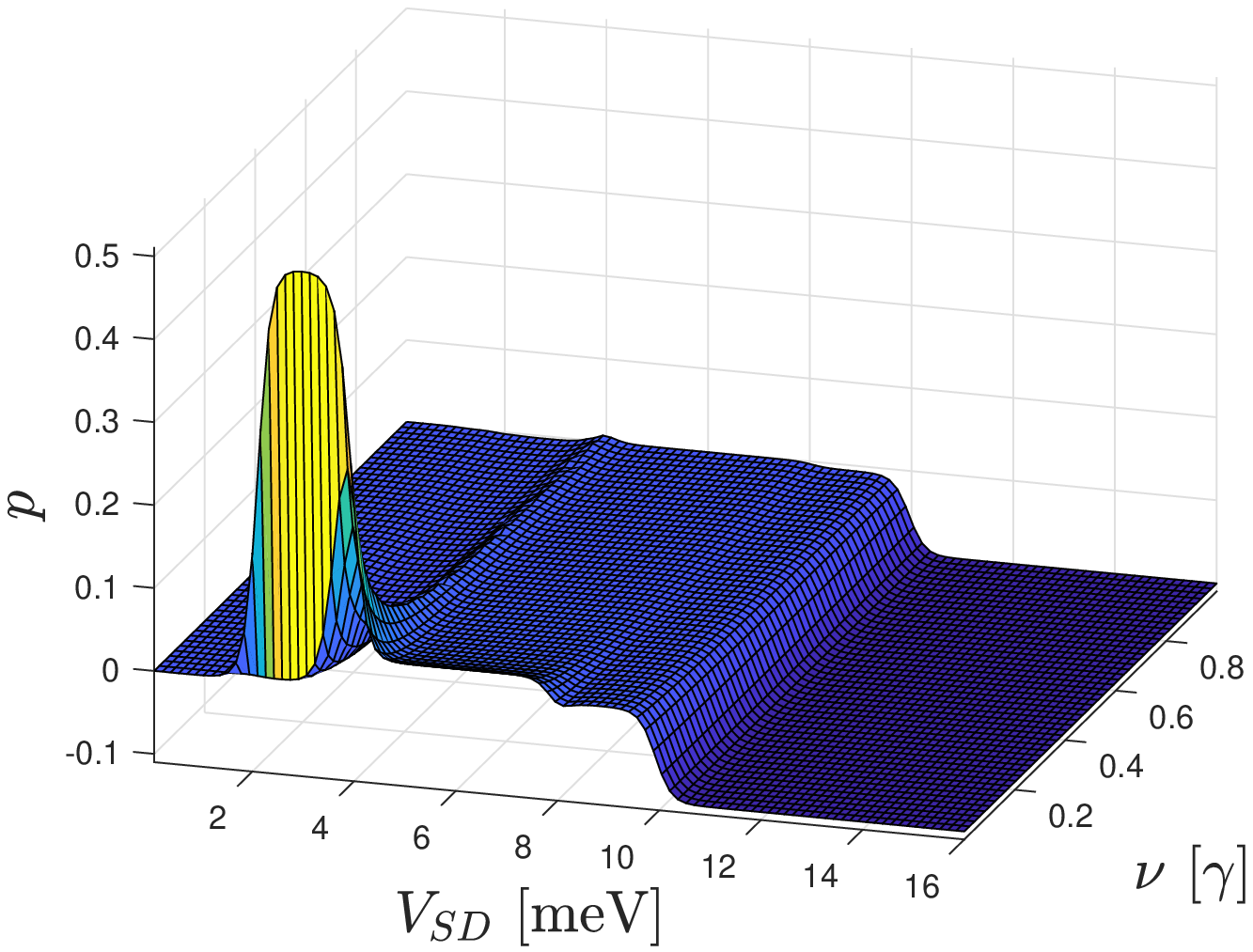}
	\caption{}	
	\label{Pearson Anderson U 1 surface}
	\end{subfigure}
    \begin{subfigure}{0.49\textwidth}
    \includegraphics[width=\textwidth]{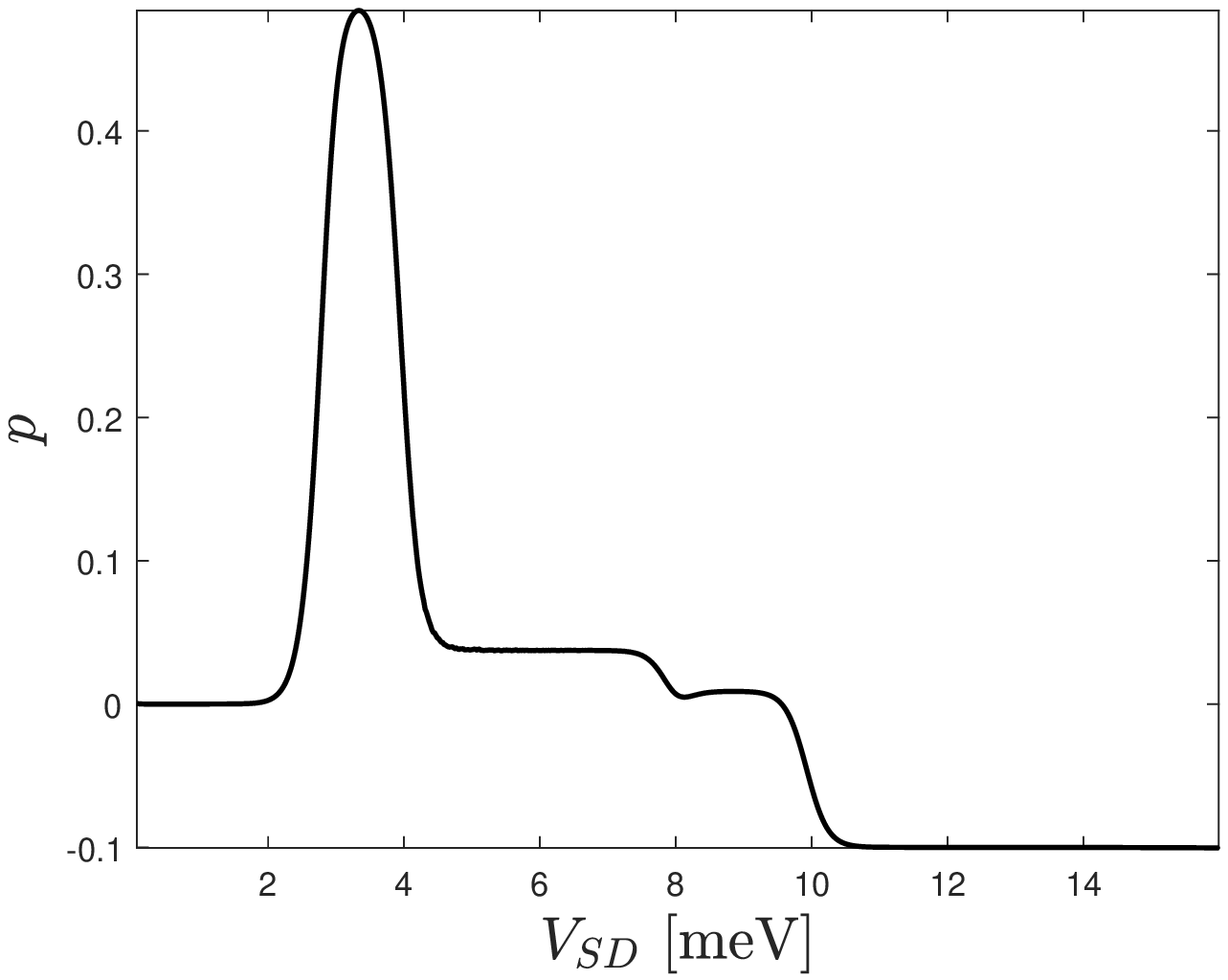}
	\caption{}	
	\label{Pearson Anderson U 1}
	\end{subfigure}
    \caption{Color online. Pearson correlation coefficient as a function of (a) $V_{SD}$ and $\nu$, and (b) as a slice at $\nu = 10^{-4}\gamma$. All other parameters are the same as in Fig.(\ref{Current Anderson U = 1}) and Fig.(\ref{Fano Anderson U = 1}).}
\end{figure*}

The Fano factor and randomness parameter diverge to an even greater degree over the same voltage, reinforcing that this is non-renewal behavior. In fact, Fig.(\ref{Fano Anderson U = 1}) shows that $F$ and $R$ differ at all voltages except $V_{SD}/2 < 1\text{meV}$, while this feature is difficult to see from the current alone. Fig.(\ref{Pearson Anderson U 1}) confirms the non-renewal behavior as $p \approx 0.5$ peaks between $1\text{meV} \leq V_{SD}/2 \leq 2\text{meV}$; as expected, there are relatively strong positive correlations accompanying telegraphic switching. What remains now is to analyze \textit{why} the characteristic first-passage times of scenario $a$ and $b$ are so different in this regime. We note that, since $F,R \rightarrow +\infty$ as $V_{SD} \rightarrow 0$, we plot all results starting just outside this region.

The degenerate $\varepsilon^{a}_{\sigma}$ level is fully open at $V_{SD}/2 = 1\text{meV}$, but all levels for the $b$ scenario remain closed. As $V_{SD}/2$ increases from $1-2\text{meV}$, the $\varepsilon^{b}_{\uparrow}$ begins to open due to thermal effects in the baths. Because it is not fully open, though, the current (first-passage time) through $\varepsilon^{b}_{\uparrow}$ in this voltage regime is much smaller (greater) than that through $\varepsilon^{a}_{\sigma}$. 

In Fig.(\ref{Pearson Anderson U 1 surface}), we have also plotted the Pearson correlation coefficient, Eq.\eqref{Pearson FPTD}, as a function of $V_{SD}$ and $\nu$. In it, we see that as $\nu$ increases the correlation in $1\text{meV}\leq V_{SD}/2 \leq 2\text{meV}$ decreases, until it is close to zero when $\nu = \gamma$. Physically, if $\nu = \gamma$, then the molecule switches between configurations at the same rate at which electrons enter and leave, so that the system does not spend long enough in either configuration for significant correlations between successive first-passage times. Also noticeable is that the correlation peak shifts closer to $V_{SD}/2 = 2\text{meV}$, since at larger $\nu$ the system does not spend long enough in the $b$ configuration to record many tunnelings, and thus correlations, at lower voltages.

Apart from identifying non-renewal behavior, the second cumulants reveal telegraphic switching behavior in the magnitude of their peaks, which are $\sim 10^{3}$. This is unusual for Markovian quantum systems, in which transport is usually close to Poissonian and $F,R \propto 1$. Such large $F$ and $R$ arise from the large differences between the characteristic first-passage times $\langle \tau \rangle_{a}\text{ and }\langle \tau \rangle_{b}$. As the voltage increases and telegraphic switching influences the transport less, these effects accordingly disappear from the $F,R$ and $p$. 

As $V_{SD}/2$ approaches $4\text{meV}$, the $\varepsilon_{\downarrow}^{b}$ begins to open and there is little difference between scenario $a$ and $b$. Between $2\text{meV} < V_{SD} < 4\text{meV}$ the noise is thus better described by the Fano factor under the Coulomb blockade \cite{Bagrets2003}, $F = \left((\gamma^{S})^{2} + 4(\gamma^{D})^{2}\right)/\left(\gamma^{S} + 2\gamma^{D}\right)^{2}$, which for symmetric coupling reduces to $F = 5/9$. Indeed, at $V_{SD}/2 = 4\text{meV}$ the Fano factor comes close to this value, although it is not visible in Fig.(\ref{Fano Anderson U = 1}). At higher voltages, when $V_{SD}/2 > 5\text{meV}$ all energy levels are open and the system is effectively non-interacting; the corresponding noise is the well-known result \cite{Nazarov1996,Chen1992} $F = \left((\gamma^{S})^{2} + (\gamma^{D})^{2}\right)/\left(\gamma^{S} + \gamma^{D}\right)^{2}$. Since we use symmetric couplings the Fano factor reduces to $F = 0.5$, which is the exact value in Fig.(\ref{Fano Anderson U = 1}). The corresponding Pearson coefficient, in this regime, is $p \approx -0.1$: the standard result for a single Anderson impurity in the high-bias limit.
\subsection{Vibrational switching: Holstein model}

\begin{figure*}
		\begin{subfigure}{0.4\textwidth}
			\centering
			\includegraphics[width=\textwidth]{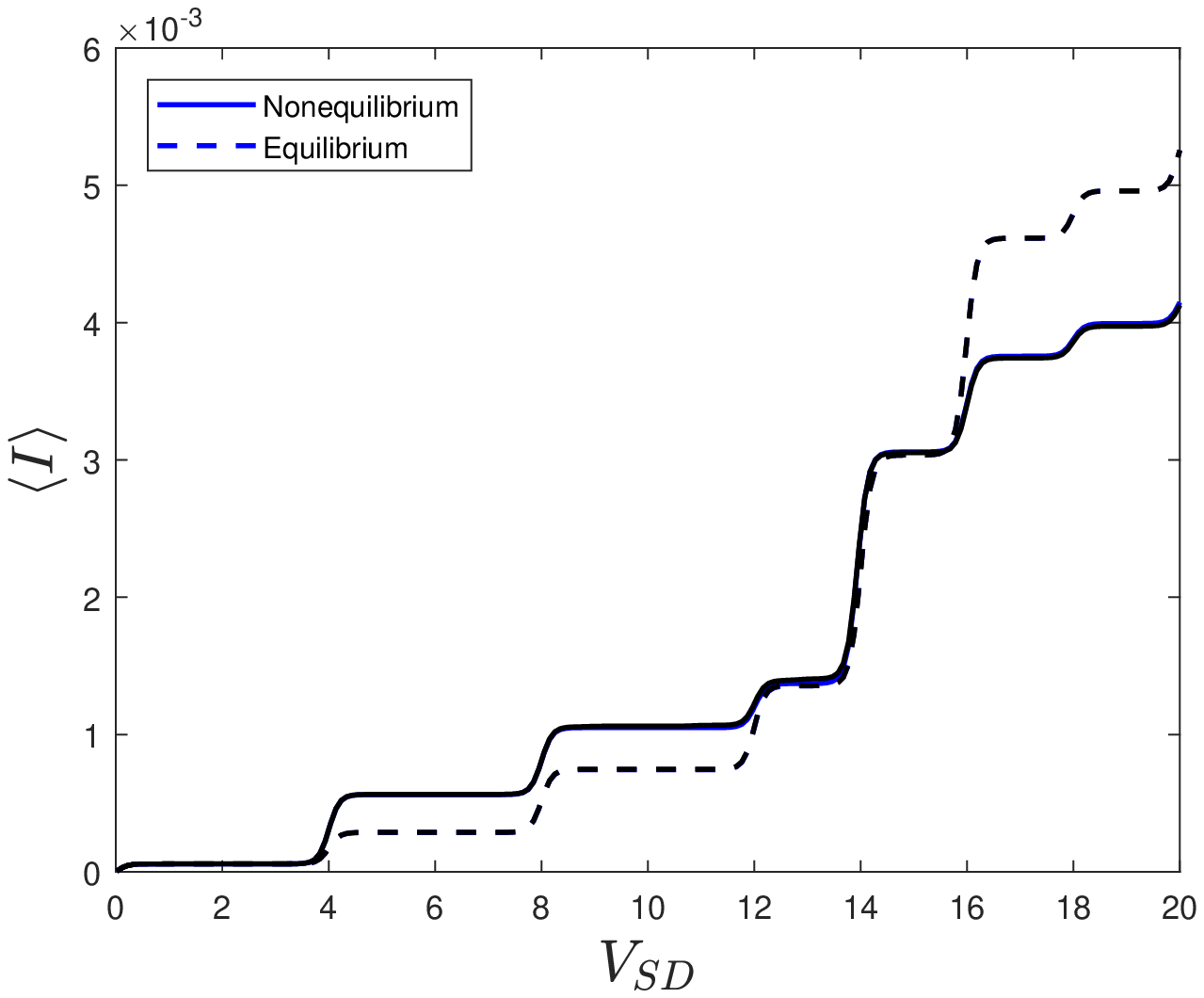}
			\caption{} 
			\label{Holstein_polaron_current}
		\end{subfigure} \: \begin{subfigure}{0.4\textwidth}
			\centering
			\includegraphics[width=\textwidth]{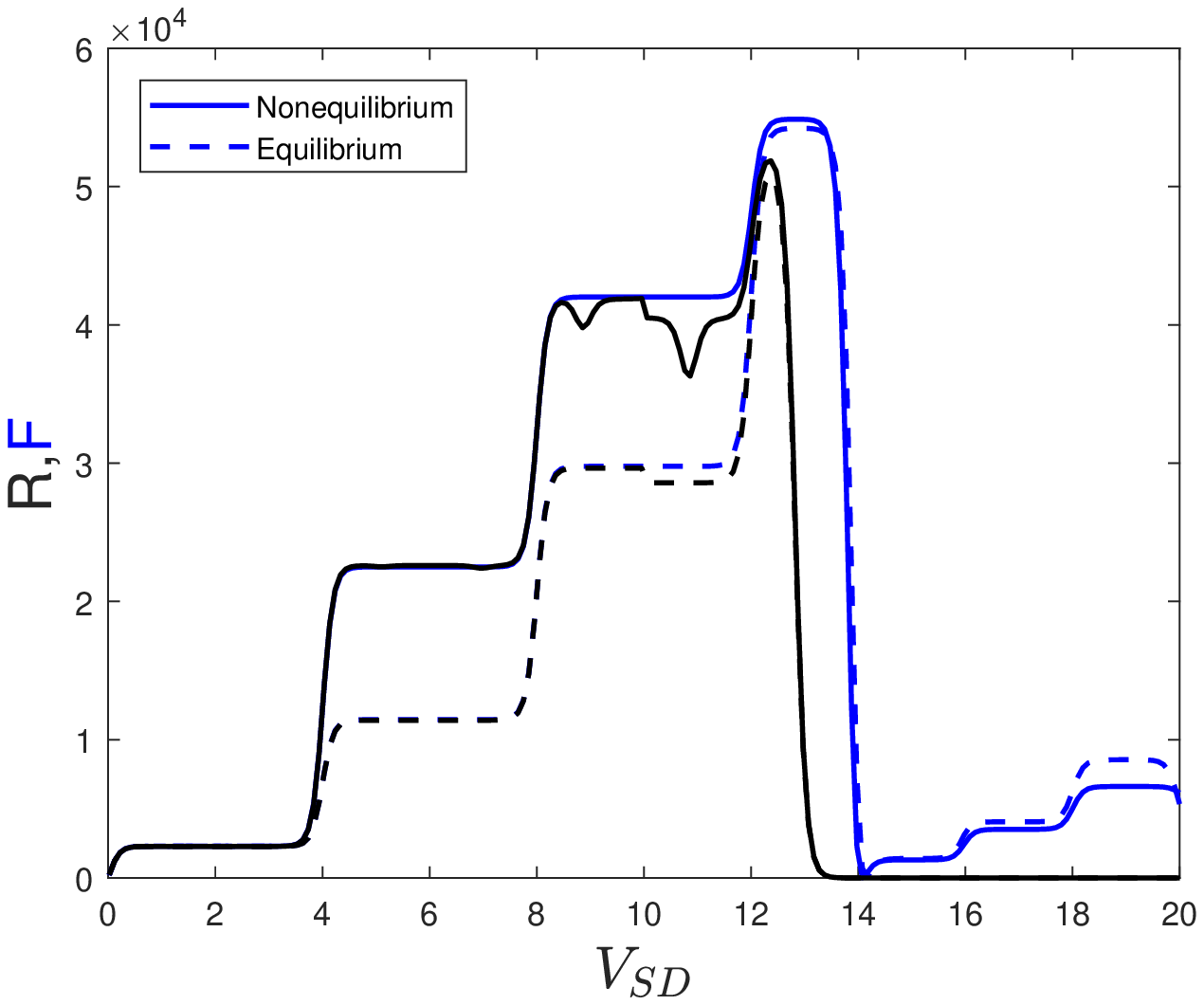} 
			\caption{}
			\label{Holstein_polaron_fano}
		\end{subfigure}
    \caption{Color online. Comparison of the first (a) and second (b) cumulants of the FCS and the FPTD as a function of $V_{SD}$, for both equilibrated and unequilibrated vibrations. The polaron shifted energy levels are $\varepsilon_{a} = \frac{\lambda_{b}^{2}}{\omega_{b}} - \frac{\lambda_{a}^{2}}{\omega_{a}}$ and $\varepsilon_{b} = 0$; the phonon frequencies are $\omega_{a} = 1$ and $\omega_{b} = 2$; and the electron-phonon couplings are $\lambda_{a} = 1$ and $\lambda_{b} = 4$. The temperature of both the source and the drain, as well as the effective phonon temperature, is $T = T_{V} = 0.05$, from which we again define $\gamma = 0.5T$. The molecule-electrode couplings are $\gamma^{\alpha,\varphi} = \frac{\gamma}{2}$ and the telegraphic switching rates are $\nu_{0} = \nu_{1} = 10^{-6}\gamma$.}
\end{figure*}

\begin{figure}
{\includegraphics[width=1\columnwidth]{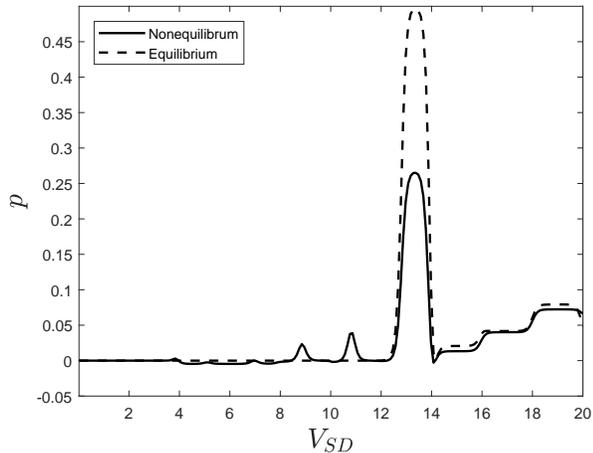}}
\caption{Pearson correlation coefficient as a function of $V_{SD}$ for the same parameters as Fig.(\ref{Holstein_polaron_current}) and Fig.(\ref{Holstein_polaron_fano}).}
\label{Holstein_polaron_pearson}
\end{figure}

In Fig.(\ref{Holstein_polaron_current}), Fig.(\ref{Holstein_polaron_fano}), and Fig.(\ref{Holstein_polaron_pearson}), we have assumed that a single level, $\varepsilon_{0}$, telegraphically switches between two different vibrational coupling configurations. The energy $\varepsilon_{0}$ is chosen as the polaron shift for configuration $b$, $\frac{\lambda_{b}^{2}}{\omega_{b}}$, so that $\varepsilon_{a} = 7$ and $\varepsilon_{b} = 0$. We note that, for all calculations from the Holstein model, we have chosen more natural units. Consequently, all energy parameters are scaled in terms of $\omega_{a}$ (or $\hbar\omega_{a}/e$ outside of natural units) and $\langle I \rangle$ is also scaled in terms of $\omega_{a}$ (or $e/\omega_{a}$).

The polaron-shifted energy of configuration $a$ is large enough that, at low voltages, many phonon interactions are required for electrons to tunnel through the molecule. For $\lambda_{a} = 1$, however, only small $|q-q'|$ transitions have non-negligible Franck-Condon matrix elements, and so the contribution to the current remains zero until $V_{SD}/2 \rightarrow 7$, when $\varepsilon_{a}$ begins to open for elastic $q=0$ transitions.

The Franck-Condon blockade \cite{Koch2005} is present at low voltages for configuration $b$, since $\lambda_{b} = 4$ and $\omega_{b} = 2$; hence the small current steps $V_{SD}/2 = 2n$. The current due to unequilibrated phonons is initially larger than that due to equilibrated phonons, since at low voltages the high $|q-q'|$ transitions required to overcome the blockade are more likely when phonons are out of equilibrium. When the voltage window contains both energy levels, however, the equilibrated phonons provide a larger current, since low $|q-q'|$ transitions dominate through the $\varepsilon_{a}$ level and transitions to the $q=0$ state are required for transport through the $\varepsilon_{b}$ level. 

Fig.(\ref{Holstein_polaron_current}), while illuminating, does not visually display any non-renewal behavior. Numerical differences between $\langle I \rangle$ and $\frac{1}{\langle \tau_{1}\rangle}$ may be present, but they are not easily visible to the naked eye. In Fig.(\ref{Holstein_polaron_fano}) and Fig.(\ref{Holstein_polaron_pearson}), however, we can easily determine the renewal behavior. When $V_{SD}/2 < 2$, $F=R$ and the transport is clearly renewal, which we can also see in Fig.(\ref{Holstein_polaron_pearson}), since in this voltage range $p = 0$. This result seems counter-intuitive: the two conformations have different characteristic currents, so why does telegraphic switching not produce positive correlations? 

The transport is renewal in this regime because the current through configuration $a$ is negligible; the telegraphic switching simply places large time gaps of no tunneling between periods of tunneling through configuration $b$. This produces ``avalanche" tunneling, in which the mean of the first-passage time distribution is much larger than the mode, and is accompanied by large Fano factors and randomness parameters $F,R \sim 10^{4}$. A similar effect occurs during extreme Franck-Condon blockades \cite{Koch2005}. 

The most noticeable non-renewal behavior is the correlation peak $p \approx 0.5$ at $V_{SD}/2 \approx 6.5$: when the $\varepsilon_{a}$ level begins to open for elastic $q=0$ transitions, which for $\lambda_{a} = 1$ are the dominant current contribution.  At this point, the current from configuration $a$ is non-negligible and telegraphic switching correlations appear. These are larger for equilibrated phonons for two reasons. First, when phonons are in equilibrium the elastic $q=0$ transition is maximized, and second, when phonons are unequilibrated, the Franck-Condon blockade is minimized. There are also two similar, but much smaller, correlation spikes at $V_{SD}/2 \approx 4.5$ and $V_{SD}/2\approx 5.5$ corresponding to $q=1$ and $q=2$ transitions beginning to open. As expected, these non-renewal regimes are accompanied by a discrepancy between $F$ and $R$.

\begin{figure*}
		\begin{subfigure}{0.4\textwidth}
			\centering
			\includegraphics[width=\textwidth]{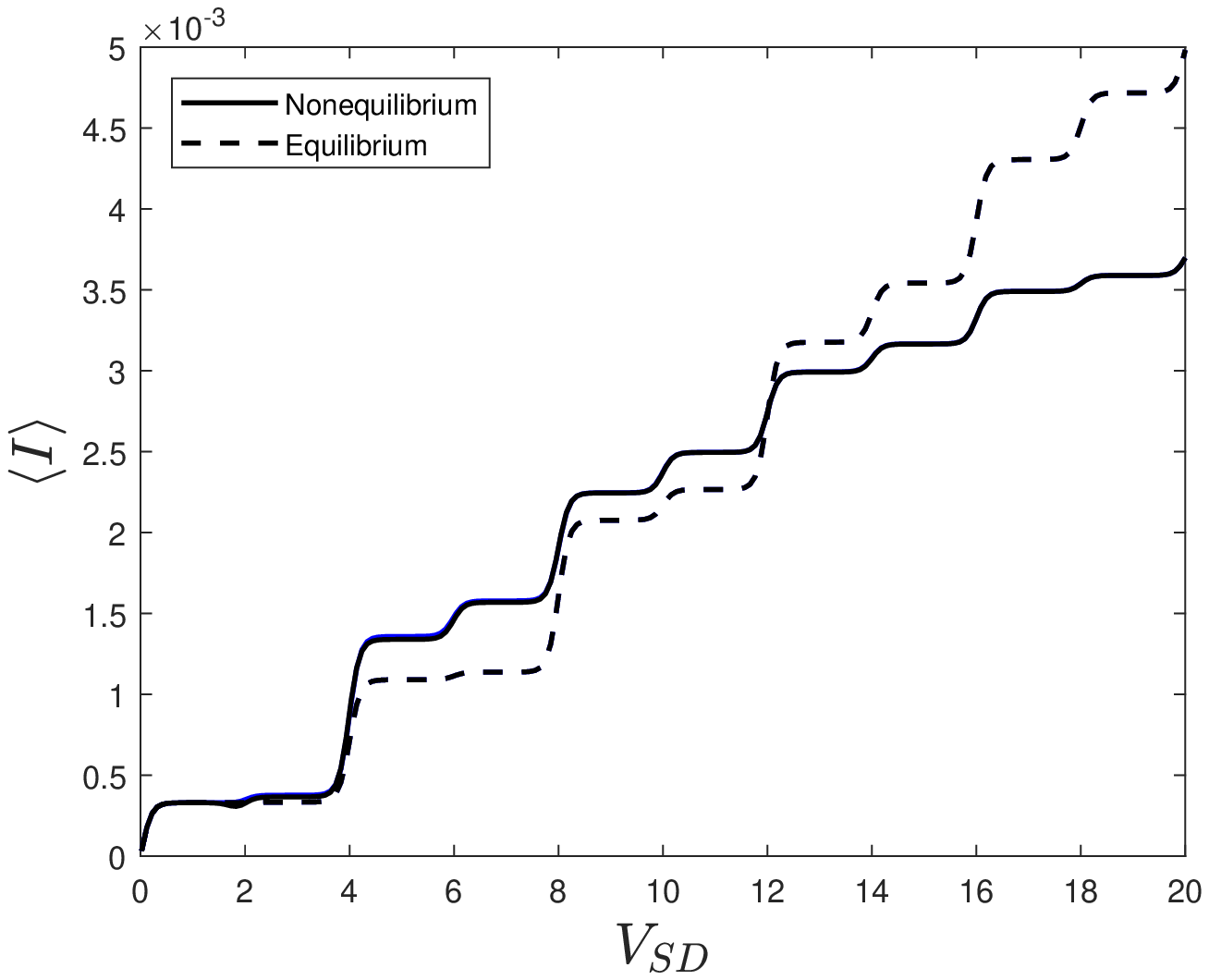}
			\caption{} 
			\label{Holstein no polaron current}
		\end{subfigure} \: \begin{subfigure}{0.4\textwidth}
			\centering
			\includegraphics[width=\textwidth]{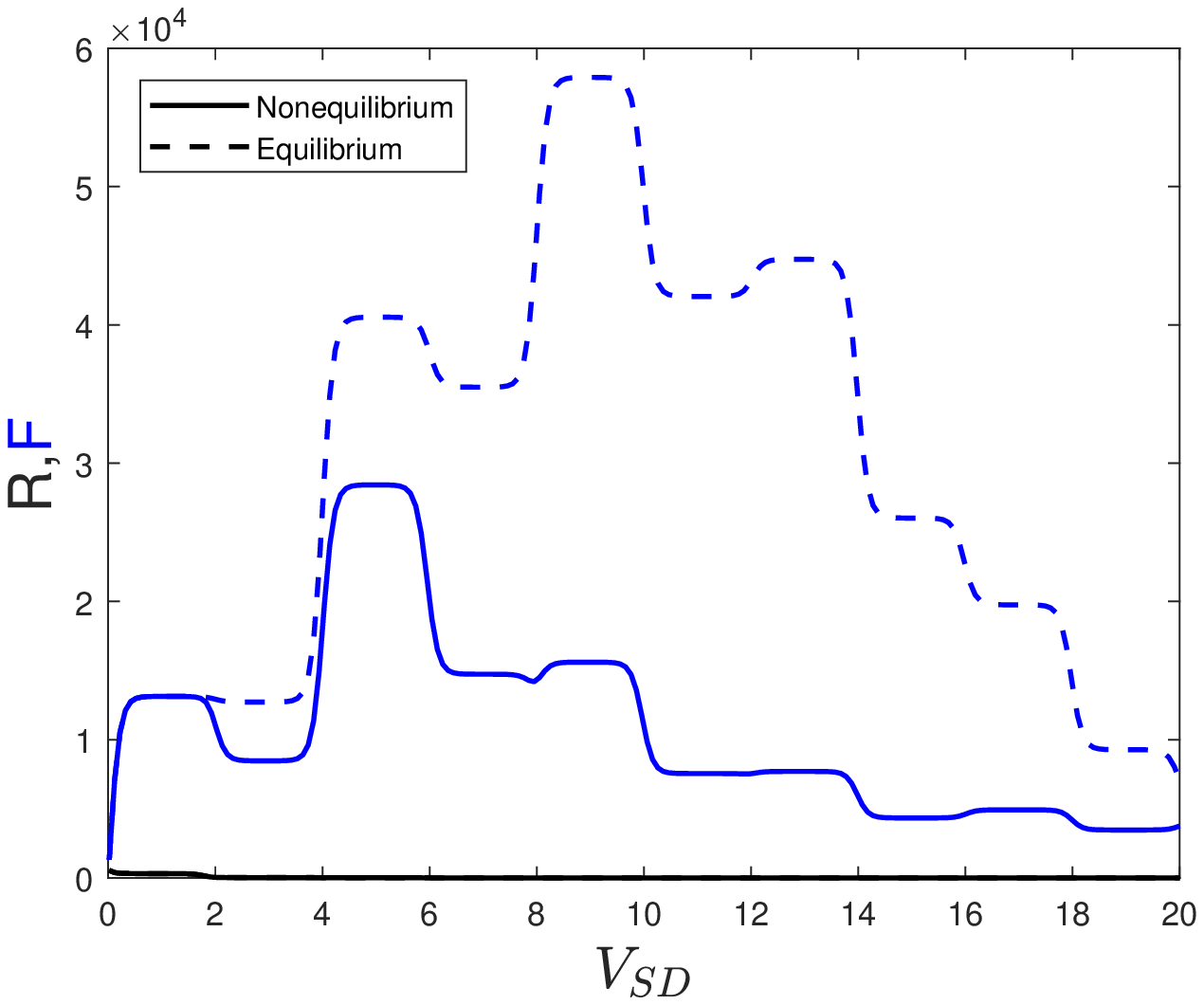} 
			\caption{}
			\label{Holstein no polaron fano}
		\end{subfigure}
    \caption{Color online. Comparison of the first (a) and second (b) cumulants of the FCS and the FPTD as a function of $V_{SD}$, for both equilibrated and unequilibrated vibrations. The polaron shifted energy levels are $\varepsilon_{a} = \varepsilon_{b} = 0$; the phonon frequencies are $\omega_{a} = 1$ and $\omega_{b} = 2$; and the electron-phonon couplings are $\lambda_{a} = \lambda_{b} = 3$.}
\end{figure*}

\begin{figure}
{\includegraphics[width=1\columnwidth]{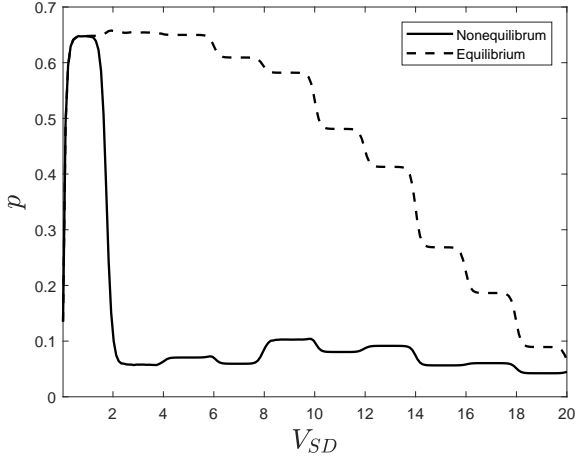}}
\caption{Pearson correlation coefficient as a function of $V_{SD}$ for the same parameters as Fig.(\ref{Holstein no polaron current}) and Fig.(\ref{Holstein no polaron fano}).}
\label{Holstein no polaron pearson}
\end{figure}

In Fig.(\ref{Holstein no polaron current}), Fig.(\ref{Holstein no polaron fano}), and Fig.(\ref{Holstein no polaron pearson}) we exclude the polaron shift by setting $\varepsilon_{a} = \varepsilon_{b} = 0$, which implies that the two molecular configurations correspond to two different orbitals separately coupled to two vibrational modes. We also keep identical electron-phonon couplings, so that telegraphic switching phenomena arises solely from the difference between $\omega_{a} = 1$ and $\omega_{b} = 2$.

At all $V_{SD}$ in Fig.(\ref{Holstein no polaron current}), we can clearly see double steps in the $I-V$ characteristics. As with the Fig.(\ref{Holstein_polaron_current}), there are small steps at $V_{SD}/2 = (2n+1)$, corresponding to phonon interactions in configuration $a$ only, and larger steps at $V_{SD}/2 = 2n$, corresponding to phonon interactions in both configurations. The Franck-Condon blockade is present at low voltages for configuration $a$, since $\lambda_{a} = 3\text{ and }\omega_{a}=1$. However, current is not suppressed through configuration $b$, as the magnitude of the blockade effect depends on the ratio $\frac{\lambda}{\omega}$.

Although, again, we cannot see any difference between $\langle I \rangle$ and $\frac{1}{\langle \tau \rangle}$, at all voltages $F \gg R$, indicating non-renewal behavior. Fig.(\ref{Holstein no polaron pearson}) corroborates this as the Pearson correlation coefficient peaks at $0.65$ between $0 < V_{SD}/2 < 1$ and never fully decays to zero. We can understand the peak in terms of the Franck-Condon blockade.

\begin{figure}
		\centering
		\begin{subfigure}{0.49\columnwidth}
			\includegraphics[width=\textwidth]{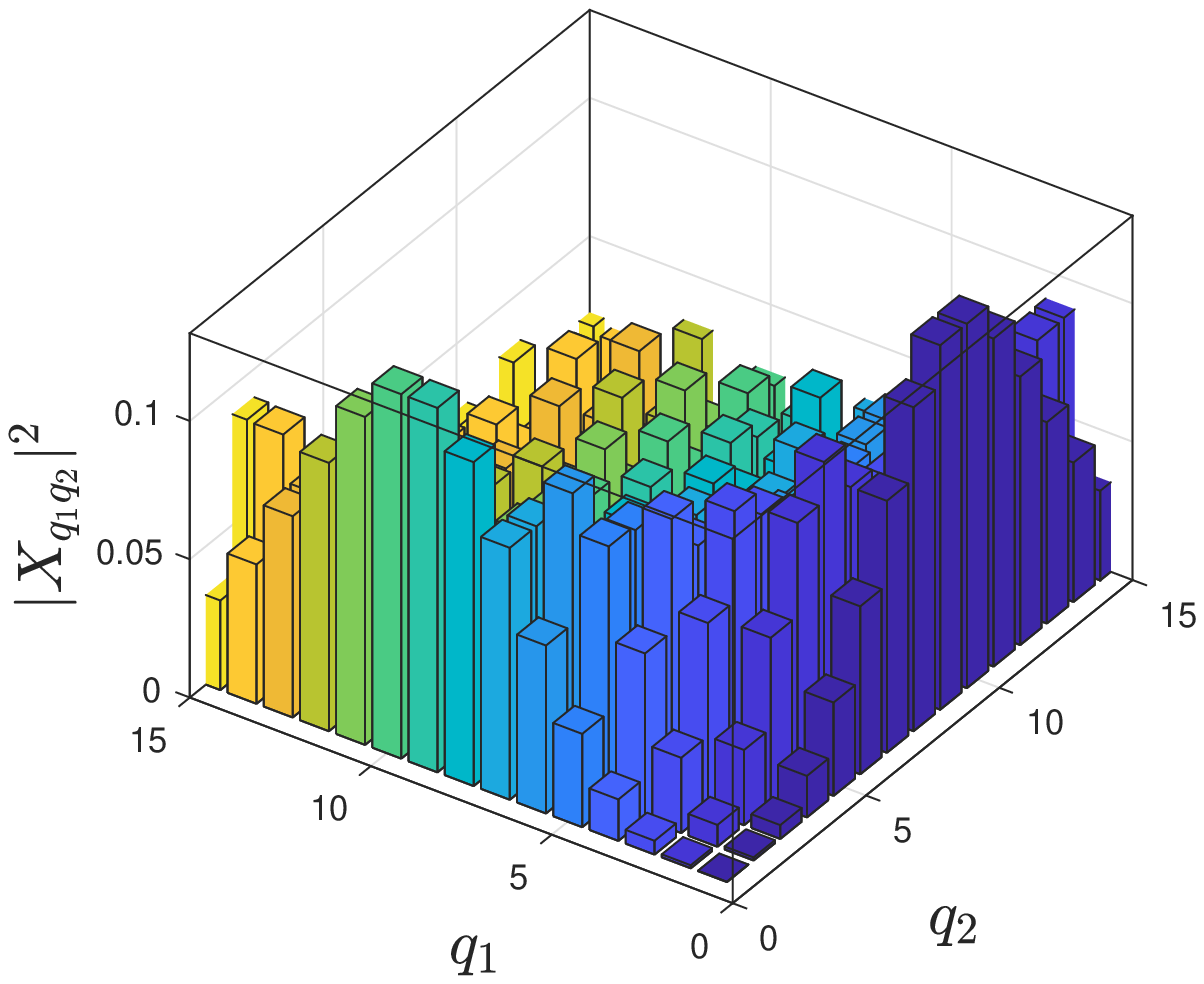}
			\centering
			\caption{} 
			\label{FC_blockade_omega1_lambda3}
		\end{subfigure} \begin{subfigure}{0.49\columnwidth}
			\includegraphics[width=\textwidth]{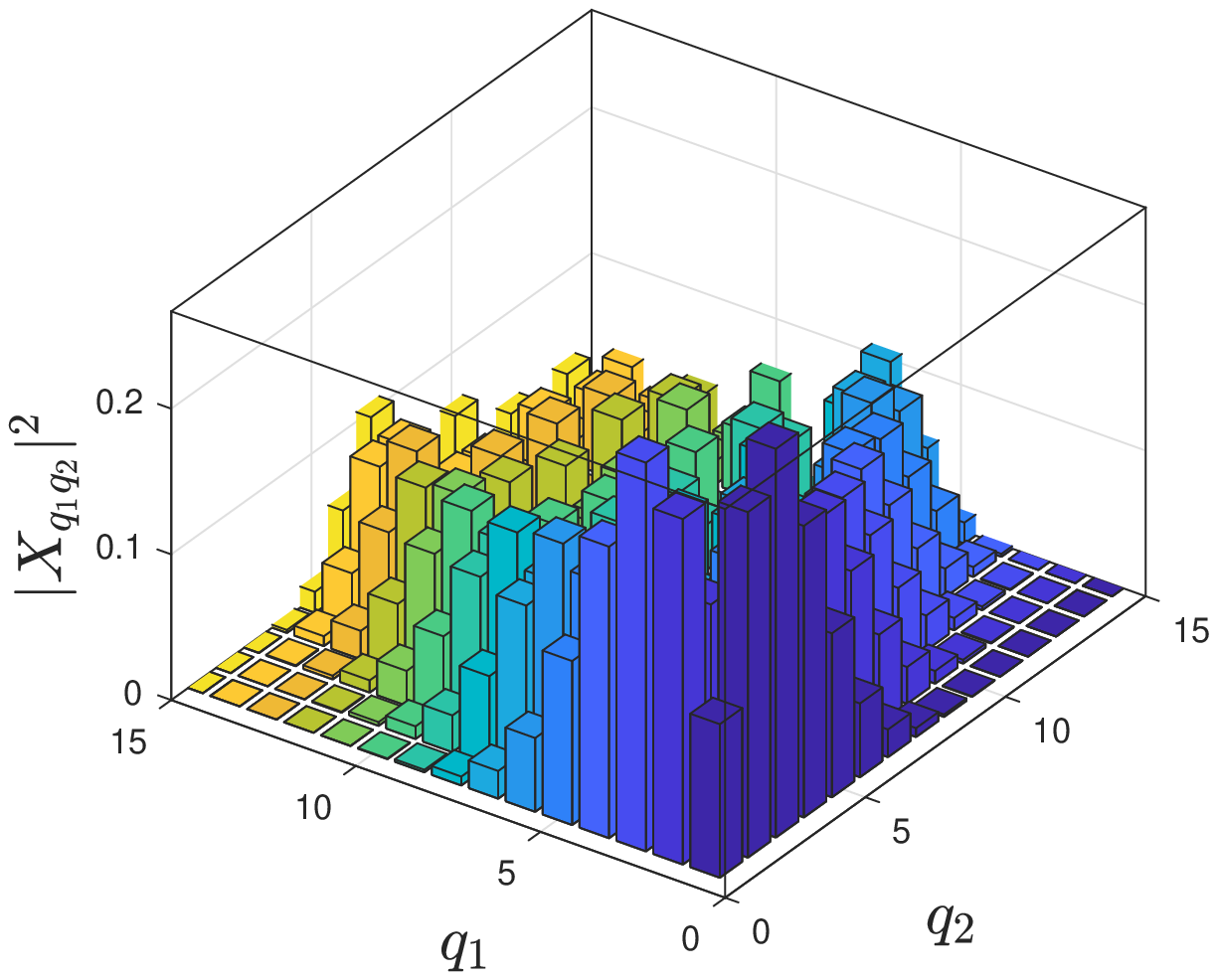}
			\centering
			\caption{}
			\label{FC_blockade_omega2_lambda3}
		\end{subfigure}
    \caption{The Franck-Condon matrix elements over a range of $q\text{ and }q'$, for two different sets of parameters: (a) $\frac{\lambda_{a}}{\omega_{a}} = 3$ and (b) $\frac{\lambda_{b}}{\omega_{b}} = \frac{3}{2}$.}
\end{figure}

The matrix elements $|X_{q_{1}q_{2}}|^{2}$ in Fig.(\ref{FC_blockade_omega1_lambda3}), for $V_{SD}/2 < 1$, are suppressed for low $|q-q'|$ and especially for $q = q' = 0$. In contrast, the matrix element $|X_{00}|^{2}$ in Fig.(\ref{FC_blockade_omega2_lambda3}) is non-zero. The elastic $q=0$ transition, therefore, is available to configuration $b$ in the voltage range $0 < V_{SD}/2 < 1$, but not to configuration $a$. Since this is the only transition available in this voltage range, there is current through configuration $b$ and no current through configuration $a$: hence the large correlations between successive first-passage times. Because the Franck-Condon blockade affects equilibrium phonons more than unequilibrated phonons, the correlations for equilibrated phonons last into higher voltages than that for unequilibrated, which decay to $p \approx 0.1$ immediately outside of $0 < V_{SD}/2 < 1$. 

\subsection{Noise on the interface: Holstein model}

\begin{figure*}
		\begin{subfigure}{0.4\textwidth}
			\centering
			\includegraphics[width=\textwidth]{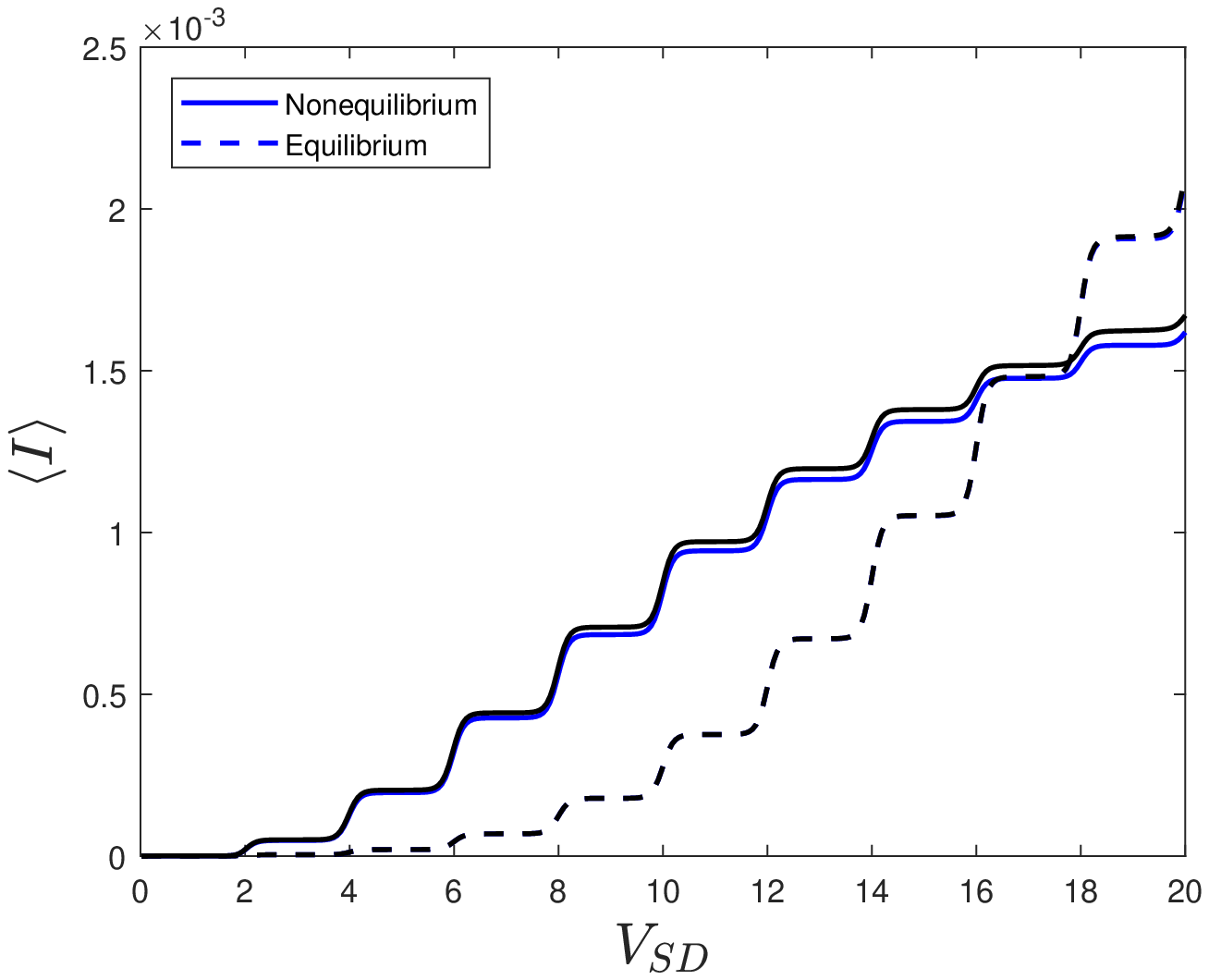}
			\caption{} 
			\label{Gamma Holstein current}
		\end{subfigure} \: \begin{subfigure}{0.4\textwidth}
			\centering
			\includegraphics[width=\textwidth]{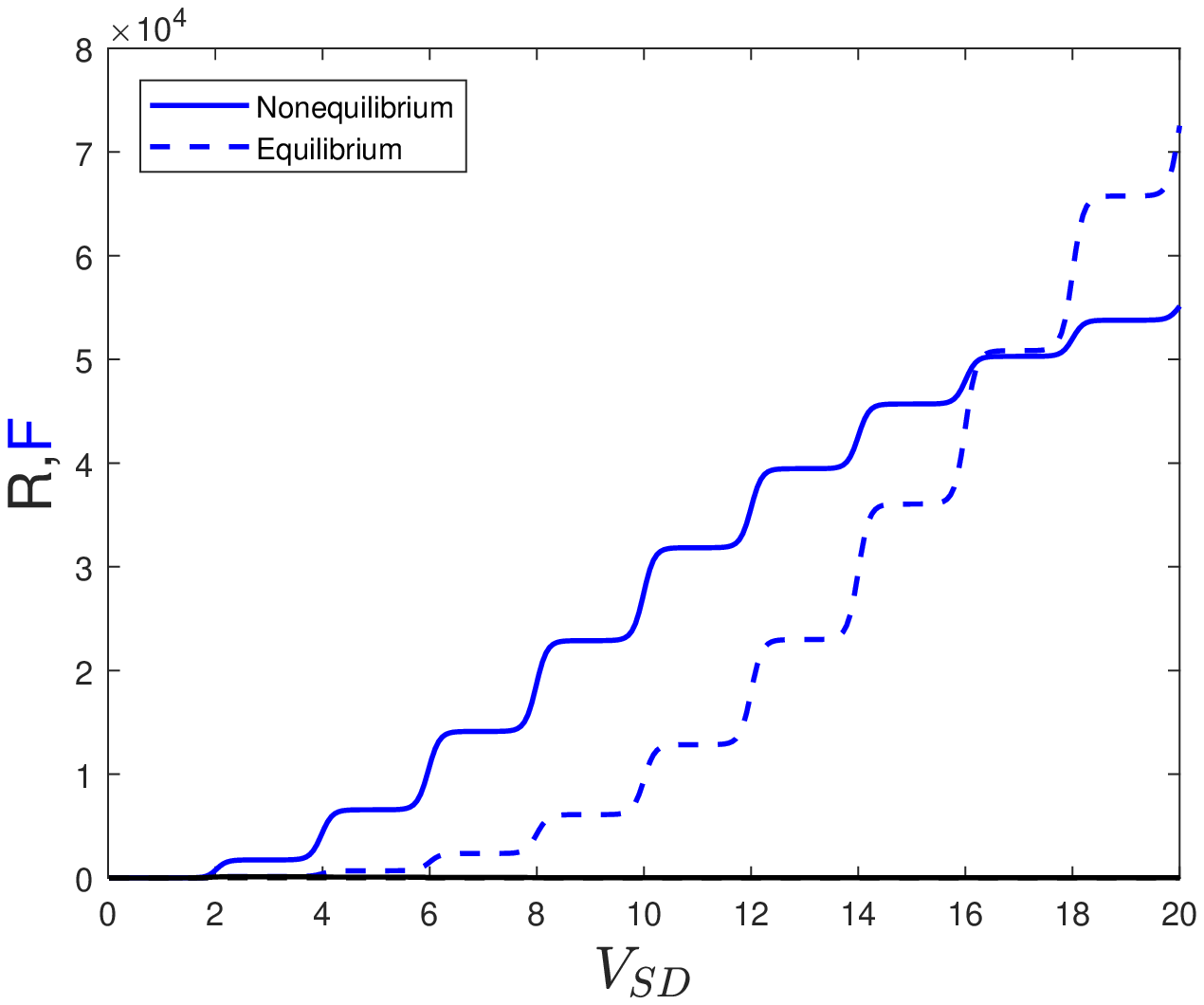} 
			\caption{}
			\label{Gamma Holstein fano}
		\end{subfigure}
    \caption{Comparison of the first (a) and second (b) cumulants of the FCS and the FPTD as a function of $V_{SD}$. The only difference between configuration $a$ and $b$ are the molecule-lead couplings. We first define a constant $\gamma = 0.5T$, and then $\gamma^{S,\varphi} = \gamma^{D,a} = \gamma/2$, and $\gamma^{D,b} = 0.01\gamma$. Else, they share the same parameters: $\varepsilon_{\varphi} = 0$, $\lambda_{\varphi} = 3$, $\omega_{\varphi} = 1$, $T = 0.05$, $T_{V} = 0.05$, $\nu_{0} = \nu_{1} = \nu = 10^{-6}\gamma$.}
\end{figure*}

Our last analysis concerns a fluctuating molecular-electrode coupling. Defining a scaling constant $\gamma = 0.5T$, we fix the molecule-source coupling at $\gamma^{S}_{\varphi} = \gamma/2$, and vary the molecule-drain coupling between configuration $a$, $\gamma^{D}_{a} = \gamma/2$, and configuration $b$, $\gamma^{D}_{b} = 0.01\gamma$. In this manner, we are able to model, albeit crudely, the molecule attaching to and detaching from the drain electrode. 

\begin{figure}
{\includegraphics[width=1\columnwidth]{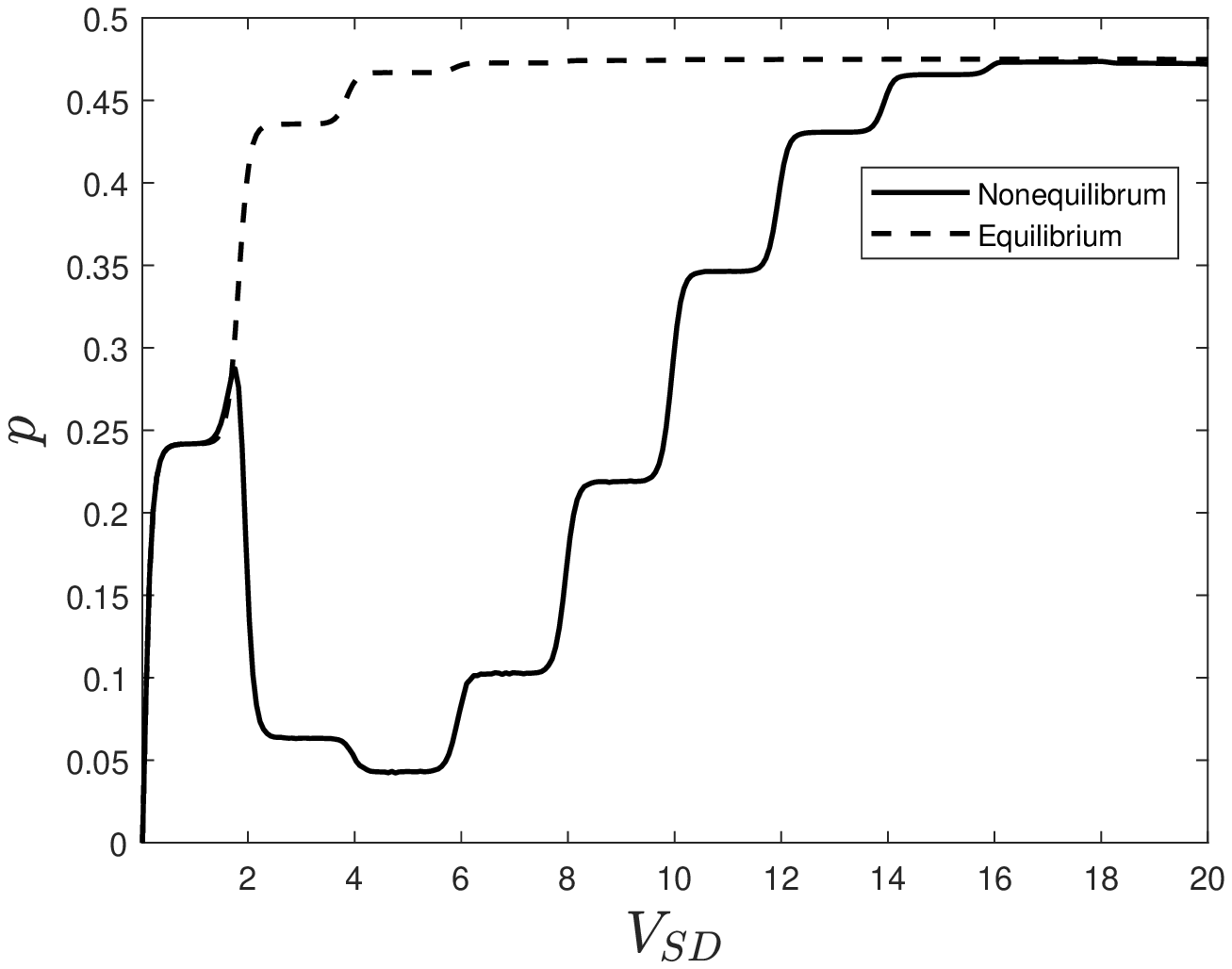}}
\caption{Corresponding Pearson correlation coefficient for the same parameters and voltage range as Fig.(\ref{Gamma Holstein current}) and Fig.(\ref{Gamma Holstein fano}).}
\label{Gamma Holstein pearson}
\end{figure}

The current associated with this process, shown in Fig.(\ref{Gamma Holstein current}), does not display double-step behavior, as $\omega_{a}=\omega_{b}$ and $\lambda_{a}=\lambda_{b}$. Fig.(\ref{Gamma Holstein fano}) and Fig.(\ref{Gamma Holstein pearson}) show that the transport dynamics are non-renewal; $F \gg R$ and $p \neq 0$, for all non-zero voltages. 

The correlation for unequilibrated phonons peaks at $p \approx 0.3$ between $0\leq V_{SD}/2 \leq 1$ before decaying to near zero and then stepping up to a maximum of $p \approx 0.5$ at higher voltages. Since $\frac{\lambda_{\varphi}}{\omega_{\varphi}} = 3$, the Franck-Condon blockade is in effect at low voltages. We surmise, then, that the weak $\gamma^{D}_{b}$ is unable to overcome the blockade and the two currents $\langle I_{a} \rangle\text{ and }\langle I_{b} \rangle$ are different enough so as to see correlations. Between $1 < V_{SD}/2 < 4$, however, configuration $a$ overcomes the blockade but $\langle I_{b} \rangle$ is still negligible. The difference between configuration $a$ and $b$ is large enough that avalanche tunneling, not telegraphic switching, is the result. At higher voltages still, $\langle I_{b} \rangle$ is now non-negligible, so telegraphically switching between the two currents $\langle I_{a}\rangle$ and $\langle I_{b} \rangle$ produces correlations. Correlations arising from equilibrated phonons, in contrast, are stable around $p \approx 0.6$ over the same voltage regime, since the Franck-Condon blockade is stronger than the difference between $\langle I_{a}\rangle$ and $\langle I _{b}\rangle$.

\section{Conclusions} \label{Conclusions}

Molecular junctions regularly undergo telegraphic switching due to a variety of physical effects. If the rate of telegraphic switching $\nu$ is much less than the rate of electron transfer $\gamma$, then the molecule spends a long time in each configuration before switching over. If the conductance difference between the two configurations $|\langle I_{a} \rangle-\langle I_{b} \rangle|$ is large, and both $\langle I_{a} \rangle\text{ and }\langle I_{b} \rangle$ are non-negligible, then successive first-passage times are positively correlated. 

Experimentally, one of the most important sources of telegraphic switching could come from an interaction with two different vibrational modes. To test this behavior, we applied the telegraphic switching rate equation to the Holstein model. We found that, when the Franck-Condon physics induced large differences between $\langle I_{a}\rangle$ and $\langle I _{b}\rangle$, there are strong positive correlations between successive first-passage times; features that are not evident from the first- and second-order current cumulants alone. The correlations, therefore, potentially provide a transport picture beyond what the current alone can see. We also found that if the current through one configuration is completely suppressed, due to the Franck-Condon blockade for example, and the other is non-negligible, then the transport is more aptly described by ``avalanche" tunneling, which is not accompanied by strong non-renewal behavior.

Via the Anderson model, we also analyzed telegraphic switching between a spin-split electronic level $\varepsilon_{\uparrow} \neq \varepsilon_{\downarrow}$ and a degenerate electronic level $\varepsilon_{\uparrow} = \varepsilon_{\downarrow}$, corresponding to stochastically switching a magnetic field $B$ on and off. We found positive correlations, with Pearson correlation coefficient $p \approx 0.5$ in voltages where the degenerate level is fully open, but only one spin-dependent level is partially open. As $\nu$ increases the correlations decrease, until they are negligible at $\nu \propto \gamma$. 

Finally, we constructed a rudimentary model of molecule-drain bonds stochastically forming and breaking, by switching between a transport scenario with normal $\gamma$, and one in which $\gamma^{D} \ll \gamma$. Here, the Franck-Condon blockade plays a role in the non-renewal behavior at low voltages. At high voltages, however, the different $\gamma^{D}$ produced strong positive correlations, which for high voltages were equal for equilibrated and unequilibrated phonons.

\section{Acknowledgments}

This work was supported by an Australian Government Research Training Program Scholarship to S.L.R. 


%

%

\end{document}